\documentclass[sigconf]{acmart}
\AtBeginDocument{%
  \providecommand\BibTeX{{%
    \normalfont B\kern-0.5em{\scshape i\kern-0.25em b}\kern-0.8em\TeX}}}

\usepackage{balance}

\newcommand{\ie}{i.e., \@}
\newcommand{\eg}{e.g., \@}

\newcommand{\etal}{et al.\xspace}
\newcommand{\eat}[1]{}

\newcommand{\curlie}[1]{Curlie}

\newcommand{\eid}{engine ID\xspace}
\newcommand{\Eid}{\expandafter\MakeUppercase \eid}
\newcommand{\eids}{{\eid}s\xspace}
\newcommand{\Eids}{{\expandafter\MakeUppercase \eid}s\xspace}
\newcommand{\etime}{engine time\xspace}
\newcommand{\Etime}{\expandafter\MakeUppercase \etime}
\newcommand{\eboots}{engine boots\xspace}
\newcommand{\Eboots}{\expandafter\MakeUppercase \eboots}
\newcommand{\lastreboot}{last reboot time\xspace}
\newcommand{\Lastreboot}{\expandafter\MakeUppercase \lastreboot}

\usepackage[normalem]{ulem}
\usepackage{savesym}
\savesymbol{Bbbk}
\usepackage{tabularx}
\newcolumntype{L}[1]{>{\raggedright\arraybackslash}p{#1}}
\newcolumntype{C}[1]{>{\centering\arraybackslash}p{#1}}
\newcolumntype{R}[1]{>{\raggedleft\arraybackslash}p{#1}}
\usepackage{graphicx}
\usepackage{subfig}
\usepackage{float}
\usepackage{subfig}
\usepackage{enumitem}
\setlist{nolistsep}
\usepackage{graphicx}
\usepackage{epstopdf}
\usepackage{grffile}
\usepackage{amssymb}
\usepackage{amsmath}
\usepackage{rotating}
\usepackage{url}
\usepackage{enumitem}
\usepackage{color}
\usepackage{xspace}
\usepackage{ifpdf}
\usepackage{mdwlist}
\usepackage{colortbl}
\usepackage{caption}
\usepackage{amssymb}
\usepackage{booktabs}
\usepackage{multicol}
\usepackage{multirow}
\usepackage{enumitem}
\usepackage{xfrac}
\usepackage{afterpage}%
\usepackage{pdflscape}
\usepackage{longtable}
\usepackage{hhline}
\usepackage{lscape}
\usepackage{pifont}
\usepackage{fancyvrb}

\usepackage{arydshln} %

\usepackage[absolute,showboxes]{textpos} %

\usepackage{hyperref}

\usepackage{cleveref}

\graphicspath{{figures/}}

\acmYear{2021}\copyrightyear{2021}
\setcopyright{rightsretained}
\acmConference[IMC '21]{ACM Internet Measurement Conference}{November 2--4, 2021}{Virtual Event, USA}
\acmBooktitle{ACM Internet Measurement Conference (IMC '21), November 2--4, 2021, Virtual Event, USA}
\acmPrice{}
\acmDOI{10.1145/3487552.3487848}
\acmISBN{978-1-4503-9129-0/21/11}

\begin{document}

\title{Measuring the Internet with SNMPv3}
\title[Third Time's Not a Charm: Exploiting SNMPv3 for Router Fingerprinting]{Third Time's Not a Charm:\\ Exploiting SNMPv3 for Router Fingerprinting}

\author{Taha Albakour}
\affiliation{%
  \institution{TU Berlin}
}
\author{Oliver Gasser}
\affiliation{%
  \institution{Max Planck Institute for Informatics}
}
\author{Robert Beverly}
\affiliation{%
  \institution{Naval Postgraduate School}
}
\author{Georgios Smaragdakis}
\affiliation{%
  \institution{TU Delft}
}

\renewcommand{\shortauthors}{Taha Albakour et al.}

\begin{abstract}
In this paper, we show that adoption of the SNMPv3 network management protocol
standard offers a unique---but likely unintended---opportunity for remotely
fingerprinting network infrastructure in the wild. Specifically,
by sending unsolicited and unauthenticated SNMPv3 requests, we obtain 
detailed information
about the configuration and status of network devices including vendor,
uptime, and the number of restarts. More importantly, the reply contains a
persistent and strong identifier that allows for lightweight Internet-scale
alias resolution and dual-stack association. By launching active
Internet-wide SNMPv3 scan campaigns, we show that our technique can fingerprint more than 
4.6 million devices of which around 350k are network routers.
Not only is our
technique lightweight and accurate, it is complementary to existing alias
resolution, dual-stack inference, and device fingerprinting approaches.  Our analysis
not only provides fresh insights into the router deployment strategies of network
operators worldwide, but also highlights potential vulnerabilities of SNMPv3 as
currently deployed.
\end{abstract}

\setlength{\TPHorizModule}{\paperwidth}
\setlength{\TPVertModule}{\paperheight}
\TPMargin{5pt}
\begin{textblock}{0.8}(0.1,0.02)
    \noindent
    \footnotesize
    If you cite this paper, please use the ACM IMC reference:
    Taha Albakour, Oliver Gasser, Robert Beverly, and Georgios Smaragdakis. 2021.
    Third Time's Not a Charm: Exploiting SNMPv3 for Router Fingerprinting.
    In \textit{ACM Internet Measurement Conference (IMC '21), November 2--4, 2021, Virtual Event, USA.}
    ACM, New York, NY, USA, 15 pages.
    \url{https://doi.org/10.1145/3487552.3487848}
\end{textblock}



\maketitle

\section{Introduction}\label{sec:intro}

Remote management functionalities are fundamental to efficient network operation.
To address this need, the Simple Network Management
Protocol (SNMP) was introduced in the 1980s and has since served as the de facto protocol
for fault notification, diagnostics, configuration management, and
statistics gathering in
IP networks~\cite{IETF-RFC1067}.
As a core IP management protocol that is widely implemented, it is unsurprising that SNMP has been
both exploited and leveraged as an attack vector---indeed, there are over 400 SNMP-related CVEs~\cite{snmp-cve}. 
The protocol itself has historically been insecure, with the first 
standardized versions (SNMPv1 and SNMPv2) including only
basic authentication via unencrypted ``community strings.'' 
Security conscious operators were therefore forced to restrict SNMP access to internal
networks.  

The current SNMPv3 standard, introduced in 2002, is implemented on
virtually all modern network equipment~\cite{IETF-RFC3411}.  The
primary focus of SNMPv3 is to provide a secure version of the
protocol by including 
mechanisms for robust authentication, integrity, and privacy. Of
direct relevance to our work is the so-called SNMP ``\eid.''
During synchronization with a client, the SNMPv3 agent exchanges
its \eid as a unique identifier.  As noted in the RFC:
the ``snmpEngineID is the unique and unambiguous identifier of an SNMP engine.
Since there is a one-to-one association between SNMP engines and SNMP entities,
it also uniquely and unambiguously identifies the SNMP
entity''~\cite{IETF-RFC3411}. 

As the \eid is integral to the protocol's key
localization mechanism,
the SNMPv3 agent
returns
this strong device identifier even in response to 
\emph{unsolicited} and \emph{unauthenticated} requests.  Moreover,
real-world implementations commonly use one of the device's
MAC addresses when forming the \eid.
This behavior offers unique---but likely unintended---opportunities for remotely
fingerprinting network devices in the wild.  We leverage the \eid
to not only identify device vendors, but also 
to perform IP alias resolution
for both IPv4 and IPv6, and thus, dual-stack identification.  
Our
method introduces a new avenue to characterize network routers
and other infrastructure that is typically impervious to traditional
methods such as TCP/IP
stack fingerprinting, 
due to closed security
postures and a lack of responding services.
And, as compared to previously explored
router identifiers such as low-entropy IP IDs (16 bits), \eids formed from MAC
addresses can provide a strong, persistent, and accurate identifier.

In addition to the \eid, 
critical information about the configuration and operation of devices
running SNMPv3 can be obtained via unauthenticated requests. For example, we show that
it is possible to retrieve the uptime of a device as
well as the number of reboots. The combination of these two parameters can also
be used as a unique device identifier.

To the best of our knowledge, previous studies and scans that considered
SNMP~\cite{Back-Door-IPv6,LZR2021} focused on SNMPv2 or SNMPv2c and were thus
unable to establish communication with the majority of devices in the
wild. In contrast,
our Internet-wide scans find more than 12 million IPs returning
a unique identifier to unsolicited and unauthenticated SNMPv3 requests.
Our controlled lab experiments, show
that some vendors enable SNMPv3 by default once SNMPv2 is enabled, suggesting
that network operators may be unaware that their devices are responding to our
queries.
Among the 12M unique SNMPv3 responses, we discover and characterize approximately 350k routers.
Our contributions thus include:

\begin{itemize}[leftmargin=*]

\item A general and lightweight technique to fingerprint the
vendor of devices running SNMPv3, including routers.

\item A new accurate and efficient large-scale alias resolution
method, including the first that can reliably identify
dual-stack IPv4/IPv6 router
aliases.  We show that our alias resolution
complements existing techniques.

\item Unique insights into router deployment
strategies by network operators worldwide, including market share
estimates and network device homogeneity.

\item Uncovering previously unrecognized
concerns in potential vulnerabilities and misuse of 
SNMPv3 as deployed in the wild.

\item Sharing our datasets and analysis scripts with interested researchers and providing updated graphs at \textbf{\url{https://snmpv3.io}}.

\end{itemize}

\section{Background}\label{sec:background}

The Simple Network Management Protocol (SNMP) is an
Internet standard
to remotely manage, configure, and monitor 
network devices.
While a wide variety of
equipment implements SNMP, some of the most common devices running
SNMP include routers, switches, and servers.  Over several decades, SNMP
has continued to mature and evolve.

\subsection{SNMP Evolution}

\begin{figure}[t]
    \centering
    \includegraphics[width=\linewidth]{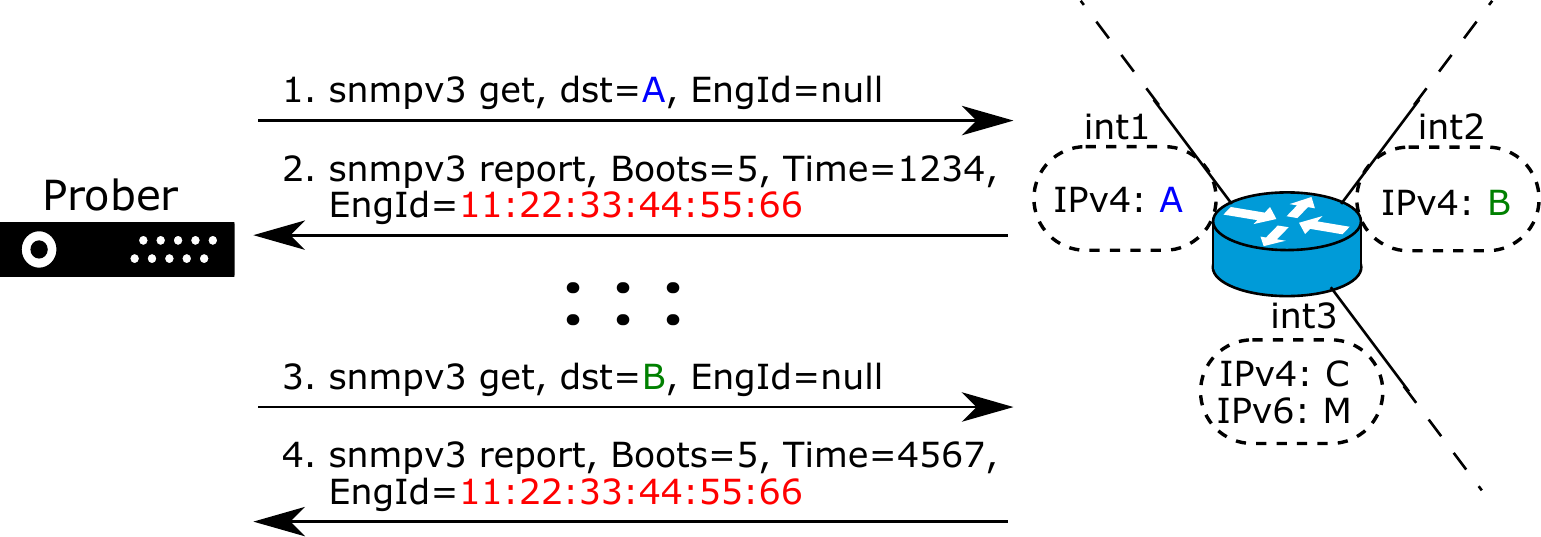}%
    \caption{Unauthenticated, unsolicited SNMPv3 probes to router IPs
             return the SNMP engine uptime, reboot count, and \eid.  The
             \eid is a unique, persistent identifier, typically formed
             from a MAC address on the device.  Here, IPv4 interfaces
             $A$ and $B$ return the
             same \eid, facilitating efficient
             alias resolution and vendor fingerprinting.  Similarly,
             our technique can identify dual-stack aliases, \eg
             IPv6 address $M$ is an alias of $A,B$ and $C$.}
    \label{fig:method} %
\end{figure}

\noindent{\bf SNMPv1:} Introduced in the late 1980s, and now deprecated,
SNMPv1 became the de facto network management
protocol~\cite{IETF-RFC1067,IETF-RFC1157}.
In SNMPv1, requests to a device are 
authenticated using a clear-text community string.  
Thus, in addition to weaknesses arising from common
and default communities or brute forcing, 
an eavesdropper able to view 
messages can easily obtain the community.
Security conscious operators were therefore forced to restrict SNMP access to internal or
out-of-band management networks.  

\noindent{\bf SNMPv2:} The second generation of SNMP
addressed some of the shortcomings of SNMPv1, 
especially performance and security~\cite{IETF-RFC1441,IETF-RFC1452}.
However, the security mechanisms of SNMPv2 were largely
not adopted due to their complexity.  Instead, the
community string-based approach of SNMPv1 was reused to create
SNMPv2c~\cite{IETF-RFC1901,IETF-RFC1908}, which was widely
deployed.
As with SNMPv1, the SNMPv2 standards are
now in historical or obsolete status within the IETF.

\noindent{\bf SNMPv3:} Our work concerns the current SNMP version~\cite{IETF-RFC3411}.  SNMPv3 was designed to be
secure by including strong user-based authentication,
integrity, replay protection, and encryption.  We detail
aspects of SNMPv3 relevant to our research next.

\subsection{SNMPv3 Unsolicited Request}

As with other SNMP versions, SNMPv3 uses UDP.
Because an explicit goal of SNMPv3 was to be stateless, it
does not employ a challenge-response authentication
mechanism~\cite{stallings}.
Instead, messages include a digest produced via a keyed HMAC (either
HMAC-MD5-96 or HMAC-SHA1-96).  This
message authentication code serves both to authenticate the message
and provide integrity protection.  In order to resist brute force
attacks and ensure that each agent stores a different key, SNMPv3
employs \emph{key localization}.  The localized key is stretched 
and derived from the user
key and the device's \emph{\eid}.  

In order to communicate with an
agent, the client must know this \eid so that it too may derive
the appropriate key to facilitate a different shared symmetric key per
device.  Thus, any client can initiate an unsolicited request, with a
missing \eid, in order to retrieve the authoritative agent's 
own \eid (steps 1 and 2 of Figure~\ref{fig:method}).  Because 
an authenticated message must, by definition, know the \eid in
order to create the HMAC, this synchronization message is not
authenticated.  Hence, \emph{any client can send this request, even
if it does not know a valid username or password.}  Note that without
valid credentials, we are still unable to retrieve any of the
device's configuration or statistics.  However, we trivially obtain
the \eid and other important meta-data including the ``\etime'' and ``\eboots''.

Compared to previous versions of the protocol, SNMPv3's authentication design makes it substantially easier to send unsolicited requests, as there is no need to guess or capture community strings.
Instead, as shown in the dissected request packet of
Figure~\ref{fig:snmpv3-request}, the synchronization request includes
an empty \eid and no password or username.
As observed by HD Moore in 2020, 
the SNMPv3 response includes, in
addition to the \eid, variables related to configuration and
operation of the device~\cite{hdmoore}. As shown in
Figure~\ref{fig:snmpv3-response}, we find information about the SNMPv3
implementation (conformance), the format of the \eid (engine ID format),
the vendor (engine enterprise ID), the number of device boots
(\eboots), and the uptime of the SNMP agent (\etime).

\subsection{SNMPv3 Unique Identifiers}

\noindent {\bf \Eid as a Unique Identifier:}  
As described in RFC 3411~\cite{IETF-RFC3411}, different information can be used to form the
\eid:
(i) an IPv4 or IPv6 address, (ii) a MAC address, (iii) ASCII text, (iv) byte values, or vendor-specific formats.

Although the \eid can be created in various ways, its value is critical to
the derived key from the key localization process~\cite{IETF-RFC3414}.
If the \eid changes, then the stored, localized keys, must be
re-generated.  Because this re-keying is cumbersome, the RFC
recommends
that the \eid ``persist across
re-initializations''~\cite{IETF-RFC3411}.

Two of the major vendors report in their documentation on the need to 
reconfigure SNMP users if the \eid
changes~\cite{Cisco-EngineID,Huawei-EngineID}.  Hence, in many popular
implementations, the \eid is not an IP address (which may
change), but instead one of the device's IEEE hardware MAC
addresses.

\begin{figure}[!bpt]
    \centering
{\scriptsize
\begin{Verbatim}[frame=single]
Simple Network Management Protocol
    msgVersion: snmpv3 (3)
    msgGlobalData
    msgAuthoritativeEngineID: <MISSING>
    msgAuthoritativeEngineBoots: 0
    msgAuthoritativeEngineTime: 0
    msgUserName: 
    msgAuthenticationParameters: <MISSING>
    msgPrivacyParameters: <MISSING>
    msgData: plaintext (0)
\end{Verbatim}
}
\caption{SNMPv3 unsolicited synchronization request.}
    \label{fig:snmpv3-request}
\end{figure}

\begin{figure}[!bpt]
    \centering
{\scriptsize
\begin{Verbatim}[frame=single]
Simple Network Management Protocol
    msgVersion: snmpv3 (3)
    msgGlobalData
    msgAuthoritativeEngineID: 800007c703748ef831db80
        1... .... = Engine ID Conformance: RFC3411 (SNMPv3)
        Engine Enterprise ID: Brocade Communication Systems, Inc.
        Engine ID Format: MAC address (3)
        Engine ID Data: BrocadeC_31:db:80 (74:8e:f8:31:db:80)
    msgAuthoritativeEngineBoots: 148
    msgAuthoritativeEngineTime: 10043812
    msgUserName: 
    msgAuthenticationParameters: <MISSING>
    msgPrivacyParameters: <MISSING>
    msgData: plaintext (0)
\end{Verbatim}
}
\caption{SNMPv3 synchronization response containing \eid.}
    \label{fig:snmpv3-response}
\end{figure}

Thus, the \eid can be considered a unique identifier of a
device: once it is generated, it is not recommended to change.  While
we do not expect collisions when devices use MAC addresses for \eid,
we acknowledge that they may use non-device unique strings, be
empty, or be ill-formatted.
We address these issues via a thorough filtering
process in 
Section~\ref{sec:datasets}.

\noindent {\bf (\Lastreboot, \Eboots) Tuple as a Unique Identifier:} 
As shown in Figure~\ref{fig:snmpv3-response}, the synchronization
request also returns the SNMP agent's \etime and \eboots. The \etime
is the number of seconds since the last time the authoritative SNMP engine has been
``booted''%
.
By subtracting the scan time we obtain the \lastreboot, \ie the time and date at which the SNMP engine was last rebooted.
The \eboots value on the other hand is the number of times the authoritative SNMP engine has
restarted. These values are included as part of the
synchronization in order to provide timeliness mechanisms, \ie to
prevent replay attacks and to detect duplicate
messages~\cite{stallings}.

It is unlikely that two devices have exactly the same
\lastreboot and \eboots. The only case that this occurs is if these
devices take an identical amount of time to boot, and 
were restarted at exactly the same point in time, \eg 
if they are co-located when a power outage occurs.
Although the
combination of (\lastreboot, \eboots) is a slightly weaker unique identifier, it can
be used in combination with the \eid to differentiate, \eg devices with the
same \eid due to misconfiguration.
We find that more than 97\% of all IPs provide (\lastreboot, \eboots) tuples belonging to a unique \eid (cf. \Cref{sec:appendix:uniqueness}).

\section{Methodology}\label{sec:methodology}

Conceptually, our methodology is straightforward.  As depicted in
Figure~\ref{fig:method}, we send unsolicited UDP SNMPv3 synchronization
packets to IPv4 or IPv6 addresses without username or password 
credentials (\eg to IP $A$ in step 1).  Note that we send well-formed,
compliant SNMPv3 packets.
If the router with the target IP address is running SNMPv3, it will
return the \eid, uptime, and boots meta-data as part of the
normal protocol synchronization (step 2).  The meta-data can
frequently be used to identify the SNMPv3 implementation and device
vendor, for instance when it is formed from a MAC address (in this
example, the \eid is \texttt{11:22:33:44:55:66}).  

This single packet
exchange process repeats for all target IPv4 and IPv6 addresses in the scan
campaign.  If a probe to an IP address returns the same unique
identifier, then we have inferred an alias.  For example, in steps 3 and 4
of Figure~\ref{fig:method}, IP $B$ returns the same MAC address as
received when $A$ was queried. Finally, by probing
both IPv4 and IPv6 addresses, we can discover dual-stack aliases,
and thus infer that IPv4 addresses $A$ and $B$ are also aliases
of IPv6 address $C$.

\subsection{Using SNMPv3 Unique Identifiers}

When both the \eid and the (\etime, \eboots) tuple are present in
the response, we have two strong unique identifiers that can be used to identify
a single device.
Further, we find that in practice, we must employ a series of filters
and tests to ensure consistent and reliable inferences.  We
describe these filtering operations in
Section~\ref{sec:datasets:engineid}.

By
collecting this information from SNMPv3 responses, we can perform a
number of measurement tasks that otherwise would have required
extensive
active probing, massive data
analysis, complex techniques, and would lack ``ground truth'' information. 
Further, our technique is complementary to existing approaches, which 
often cannot obtain a usable response from a target,
such that it increases overall inference coverage.

\noindent{\bf SNMPv3-based Alias Resolution:} Router interface IPs that are
associated with the same unique identifier are mapped to the same router. 

\noindent{\bf SNMPv3-based Dual-stack Inference:} Router interface IPs of different
protocols (IPv4 and IPv6) that are associated with the same unique identifier
are mapped to the same router.

\noindent{\bf SNMPv3-based Vendor Fingerprinting:} The information contained in the \eid can
be used to infer the vendor of the router.  Our confidence in the
inference of device vendor is highest when the \eid is generated using
a MAC address, as the upper three bytes encode the IEEE Organizational
Unique Identifier (OUI), \ie the company that registered a particular
block of MAC addresses.  Nevertheless,
useful information about the vendor can be retrieved if the \eid is
generated differently.  The Enterprise ID, that encodes the
manufacturer of the device, is always present in the SNMPv3 response if RFC
3411 is followed, and can provide more confidence or be used as an alternative when
the \eid does not unveil the vendor. 

\noindent{\bf SNMPv3-based Uptime:} In addition, the \etime value allows us to determine the uptime of routers.
This can be valuable information in order to determine, \eg the patch
status of devices, network reliability, and outage statistics.

\subsection{SNMPv3 Active Scanning Campaigns}

As there are no Internet-wide SNMPv3 scan results available, we perform our own active
scanning campaigns.
We use ZMap~\cite{ZMap} to initiate SNMPv3 unsolicited
synchronization requests by sending an SNMPv3 payload to UDP port 161.
During the scans, we capture all
SNMPv3 synchronization replies.  We perform our IPv4 probing from a
single server in a well-connected European data center and send at a rate of 5 kpps.
For IPv6 we probe from a server located in a research network at a rate of 20 kpps.

\subsection{Ethical Considerations}

In designing our active scanning, we endeavored to minimize
any potential ethical implications or harm from our study.
First, the packets we send are not only well-formed and conforming
to the SNMPv3 protocol, they are ``normal'' packets that any
SNMP agent would expect in the course of its operation.  Our
randomized probing spreads the load and each IP receives at most
one SNMP packet.  Thus,
we have no reason to believe that our packets would impair SNMP
agents.  
Our probes are connectionless UDP packets, which are generally more
innocuous as compared to TCP packets, and greatly reduce the potential
for unintended issues related to maintaining state, \eg by firewalls or
middleboxes.
Second, we coordinated with our local network administrators
to ensure that our scanning did not harm the local or upstream
network.

Next, we follow active scanning best practices \cite{ZMap,partridge2016ethical,dittrich2012menlo} and
ensure that our prober's IP address has a meaningful DNS PTR
record, and we run a web server %
with experiment and opt-out information.
To date, we
have not received any complaints or opt-out requests. 
This is in striking contrast to
other active scan campaigns performed in the past, \eg
TLS scanning and Web port scanning, which received complaints
from the target IP owners and system administrators of the
data centers hosting the scanning infrastructure. 
We observe that some scanning targets reply with a large number of responses (cf. \Cref{sec:discussion}).
Finally, our work uncovers potentially sensitive security, robustness,
and business information about network providers.  We therefore
aggregate and anonymize our results so as to not identify any
individual network.
We publish regularly updated graphs of aggregated results at \textbf{\url{https://snmpv3.io}}.

\subsection{Limitations}

While our technique provides a new method for fingerprinting
and alias resolution, as well as affording novel insights into
operational network deployments, we do note several limitations of
our method and study.

First, in this work, we limit our study to network routers
in order to provide meaningful comparisons with existing 
topology datasets and alias resolution methods.  However, in
our Internet-wide IPv4 and IPv6 scans we obtain a large number of 
responses from SNMP agents that cannot be matched to known 
router alias sets.  These responses may represent routers 
missing in CAIDA or RIPE topologies, or may be servers, 
Customer Premise Equipment (CPE), or other devices, \eg IoT.  In
future work, we plan to investigate these responses and devices
in more detail. 

Second, a deployed router may not be configured to enable SNMPv3, or may 
block outside queries by IP-level or other access control
mechanisms.  In such instances, our technique naturally cannot provide
any vendor or alias inferences.  However, because it uses a distinct
and previously unused identifier, our method is complementary to
existing approaches.

Third, while we seek to measure core network routers, our data can
capture edge and periphery devices, especially in IPv6 where 
residential devices are a routed hop~\cite{pam20edgy}.  In such cases, the IP
address of these devices can change on time-scales shorter than
our scans~\cite{gasser2018clusters,imc21scent}, causing different \eids to be returned from the
same IP address.  While we filter these instances of inconsistency
from our data, we plan to investigate this effect in more detail
in future work.

\section{Scanning for SNMPv3 Devices}\label{sec:datasets}

SNMP has been a popular protocol for decades and, thus, has been the subject of many
measurement studies~\cite{Back-Door-IPv6,LZR2021,Router-outages-SIGCOMM2017,RTG}.
However, this prior measurement work targets SNMPv2.
In addition, Censys~\cite{Censys} and Shadowserver~\cite{shadowserver-snmp} report regular Internet-wide scanning results, but solely for SNMPv2.
Moreover, compared to our SNMPv3 measurements both services report substantially lower responsive IPv4 addresses, with 1.6M for Censys and 1.2M for Shadowserver, respectively.

\subsection{Active Scan Targets}

\begin{table*}[t]
{\small
\begin{tabular}{llrrrr}
\toprule
Measurement & Date & \#IPs & \#\Eids & \#IPs w/ valid \eid & \#IPs w/ valid \eid \& \etime \\
\midrule
IPv4 scan 1 & Apr. 16--20, 2021 & 31.8M & 18.8M & \multirow{2}{*}{{\LARGE\}}} \multirow{2}{*}{27.0M} & \multirow{2}{*}{{\LARGE\}}} \multirow{2}{*}{12.5M} \\
IPv4 scan 2 & Apr. 22--27, 2021 & 31.5M & 18.6M & & \\

IPv6 scan 1 & Apr. 13, 2021 & 182k & 68k & \multirow{2}{*}{{\LARGE\}}} \hspace{.2em} \multirow{2}{*}{152k} & \multirow{2}{*}{{\LARGE\}}} \hspace{.2em} \multirow{2}{*}{140k} \\
IPv6 scan 2 & Apr. 14, 2021 & 180k & 67k & & \\
\bottomrule
\end{tabular}
}
\caption{Overview of our SNMPv3 measurement campaigns: Number of unique SNMPv3-responsive IPs, number of unique \eids, number of IPs with valid (\ie consistent and non-filtered) \eid, \eid and \etime values, respectively.}
\label{table:campaigns}
\end{table*}

We perform active scans for SNMPv3 in IPv4 as well as IPv6.
\Cref{table:campaigns}
shows an overview of our measurements.
In addition to our own measurements, we utilize multiple
third-party datasets containing known router IP addresses and
aliases as 
shown in \Cref{table:router-datasets}.

\subsubsection{IPv4 \& IPv6 SNMPv3 Scans}

We launch two Internet-wide SNMPv3 campaigns for the IPv4 protocol in
April 2021.
By employing two measurements instead of a single one, we can filter out ephemeral addresses (cf. \Cref{sec:datasets:filtering}).
We target all \textasciitilde 2.9 Billion routable IPv4 addresses.
We receive valid SNMPv3 responses from about 31M IPv4 addresses.
For each IPv4 address, we send one packet with size 88 bytes. For the large
majority of responses, we receive one packet\footnote{We receive multiple packets
for about 0.6\% of the responding IPv4 addresses, see 
Section~\ref{sec:discussion} for a discussion.} with an average size of 130 bytes.

For IPv6, we target \textasciitilde 364M addresses in non-aliased IPv6 prefixes \cite{gasser2018clusters} of the IPv6 Hitlist Service~\cite{v6hl}.
We run two consecutive scans on April 13 and 14, 2021.
In contrast to IPv4 we receive only about 180k SNMPv3 responses from these IPv6 scans.
For each IPv6 address, we send one packet with size 108 bytes and receive a response of on average 150 bytes.

\subsubsection{Router Interface Tagging}
\label{sec:datasets:subsub:router}

\begin{table}[t]
{\small
\resizebox{\columnwidth}{!}{
\begin{tabular}{llrr}
\toprule
Router dataset & Date & IPv4 addrs. (SNMPv3) & IPv6 addrs. (SNMPv3) \\

\midrule

ITDK & Mar. 2021 & 2.9M (447k) & 533k (36k) \\

RIPE Atlas & Apr. 2021 & 560k (85k) & 260k (36k) \\

IPv6 Hitlist & 2020--2021 & n/a & 63.7M (54k) \\

\midrule

Union & & 3.1M (461k) & 65M (78k) \\
\bottomrule
\end{tabular}}}
\caption{Overview of router datasets: Number of unique router IP addresses and coverage in our SNMPv3 measurements for IPv4 and IPv6, respectively.}
\label{table:router-datasets}	
\end{table}

To annotate IP addresses belonging to router interfaces, we use publicly available
datasets, namely CAIDA's ITDK~\cite{itdk}
and intermediate hop IPs extracted from RIPE Atlas traceroute measurements \cite{RIPE-Atlas} for IPv4 and IPv6.
The ITDK is a curated dataset that also includes IPv4 router-level
topologies, as inferred by
MIDAR~\cite{keys13midar} and IPv6 router-level topologies inferred via
the Speedtrap technique~\cite{Speedtrap}.
For IPv6 only we use router addresses obtained from the IPv6 Hitlist Service traceroutes \cite{v6hl}.
In \Cref{table:router-datasets} we show an overview of these router
datasets as well as the number of matches to responses in our 
more general Internet-wide SNMPv3 scans described previously.

In IPv4 the ITDK dataset contains 2.9M router IPs with 447k of those
responsive to our SNMPv3 measurements.
RIPE Atlas adds a few thousand additionally tagged router IPs, bringing 
our total known SNMPv3 responsive router IPs to 461k.

In IPv6 we find that ITDK Speedtrap and RIPE Atlas traceroutes each cover about 36k SNMPv3 addresses.
The vast corpus of IPv6 router addresses from the IPv6 Hitlist Service contains many 
Customer Premise Equipment (CPE) device addresses which are frequently changing \cite{gasser2018clusters}, thus leading to this large size.
With this dataset we obtain the highest SNMPv3 overlap with more than 54k IPv6 addresses.
The union of all IPv6 router addresses leads to an overlap of more than 78k SNMPv3 addresses, \ie more than half of all addresses with either valid \eid or \etime (see below).

\subsection{\texorpdfstring{\Eid}{\eid}}
\label{sec:datasets:engineid}

\begin{figure}[b]
    \centering
    \includegraphics[width=\linewidth]{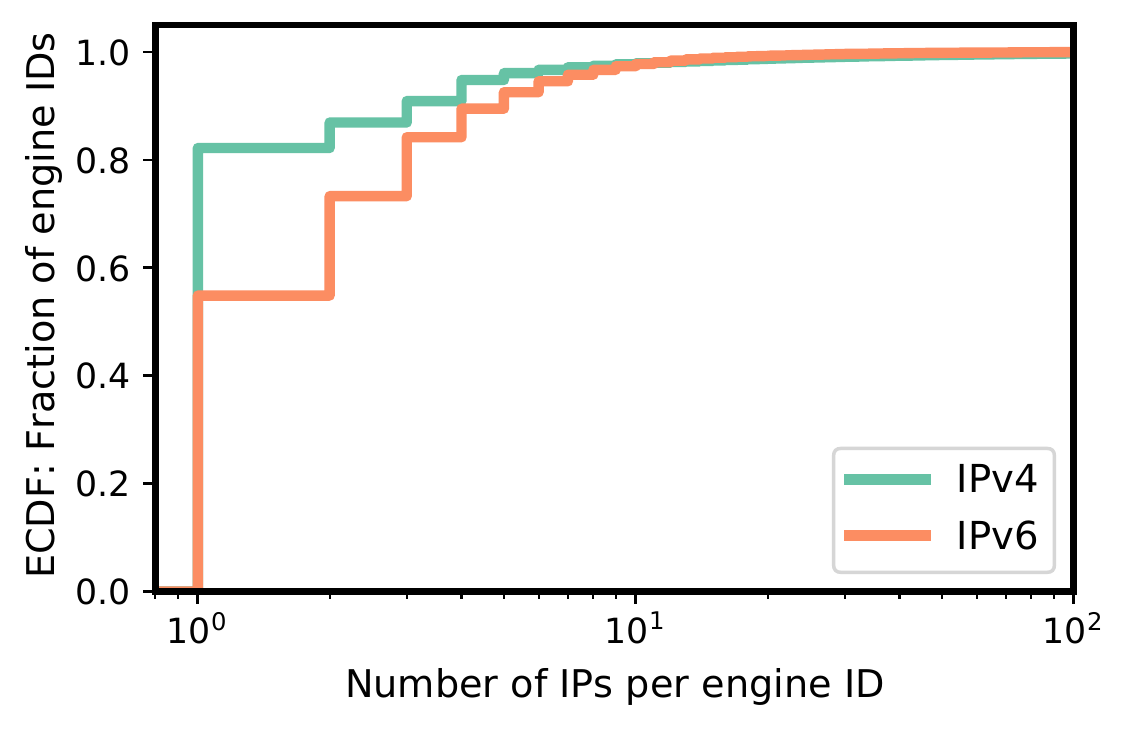}
\caption{Number of occurrences per \eid.}
    \label{fig:engineid-occurences} %
\end{figure}

To identify an SNMP agent, the protocol uses the \eid as defined by RFC 3411~\cite{IETF-RFC3411}.
In our measurements we collect 18.8M different \eids for IPv4 and 68k for IPv6.
Many SNMP devices have more than one IP address assigned to them (\eg because they are routers).
As such we see the same \eid for different IP addresses.
\Cref{fig:engineid-occurences} shows the distribution of the number of IP addresses per \eid for IPv4 and IPv6.
In IPv4 more than 80\% of \eids are seen on a single IP address.
The same is true for more than half of all \eids collected with our IPv6 measurements.
We find that the distribution is heavy-tailed, with the vast majority of 
all \eids being seen on 10 or fewer IPs, with some outliers with a
single \eid for more than 1000 IPs (not shown).

\begin{figure}[t]
    \centering
    \includegraphics[width=\linewidth]{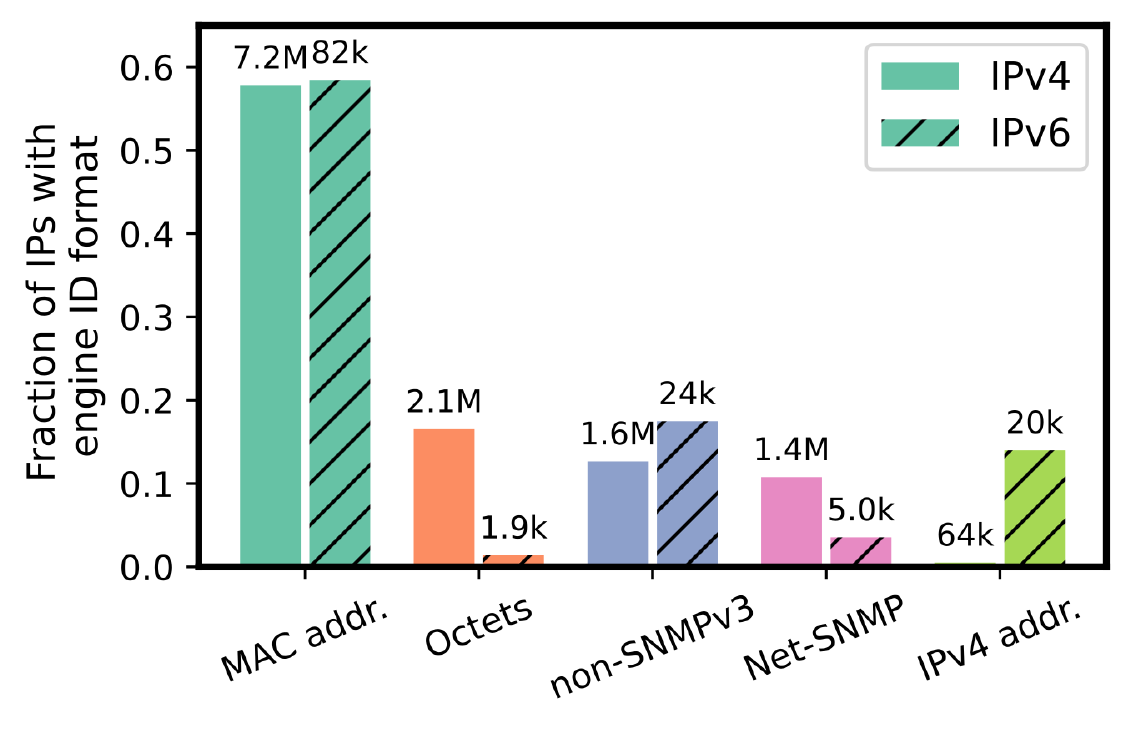}
    \caption{Distribution of different \eid formats for IPv4 and IPv6 scans.}
    \label{fig:engineid-formats}
\end{figure}

The \eid can be in different formats, \eg a device's MAC address, IP address, a text string, a byte string, or a vendor-defined format.
In \Cref{fig:engineid-formats} we show the distribution of different \eid formats.
Almost 60\% of \eids for IPv4 and IPv6 are MAC-based.
The use of MAC addresses provides a strong unique identifier and the high share of this \eid format is therefore crucial for the uniqueness of \eids.
In IPv4,
opaque byte strings (``Octets''), non-SNMPv3-conforming, and Net-SNMP-specific
\eid formats contribute 10-20\% each.
The non-SNMPv3-conforming \eids do not contain any format information but rather just the byte values (\eg \texttt{0x0300e0acf1325a88}).
Similarly, the SNMPv3-conforming Octets format also contains raw bytes values (\eg \texttt{0x3910910680002970}).
Net-SNMP on the other hand, is a popular software-based SNMP implementation \cite{netsnmp}, which uses an enterprise-specific format (\eg \texttt{0x0f010e3732bed25e00000000}).
For IPv6 we also find a relatively high share of non-SNMPv3-conforming \eids, and a rather low number of Octets and Net-SNMP formats.
Interestingly, we find more than 15\% of \eids collected in our IPv6 measurements contain IPv4 addresses.
These might reveal IPv4-IPv6 dual-stack deployments, which we investigate in detail in \Cref{sec:alias}.

\begin{figure}[!bpt]
    \centering
    \includegraphics[width=\linewidth]{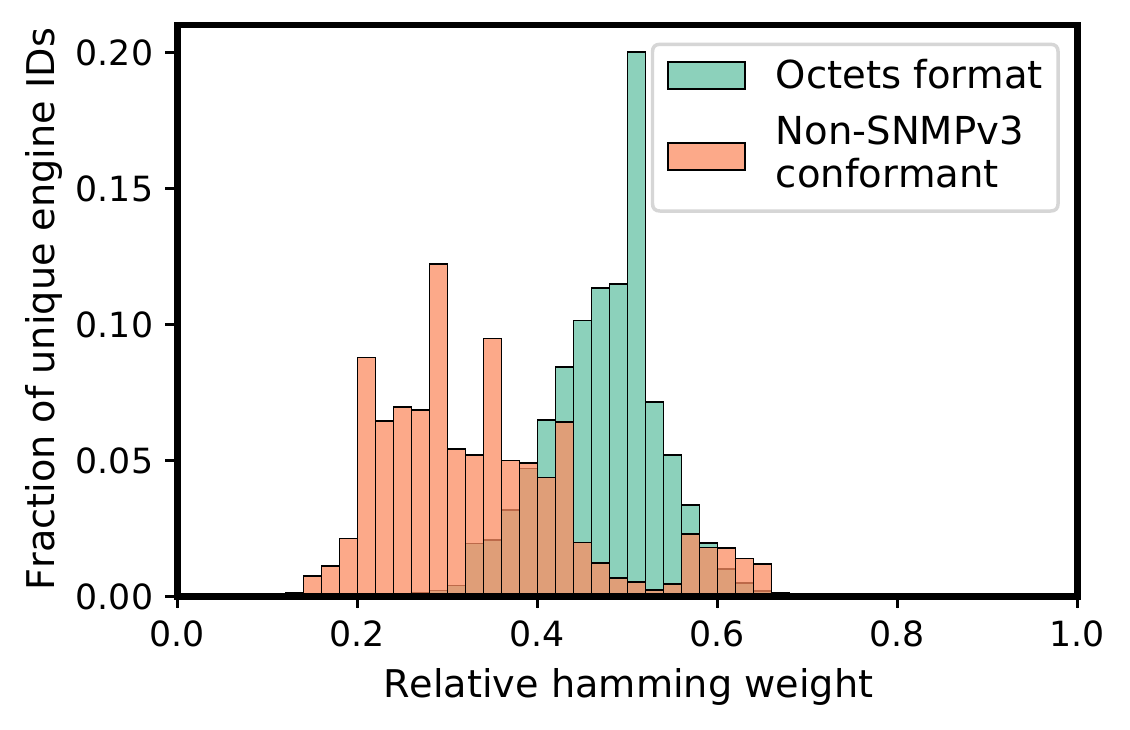}
    \caption{Relative Hamming weight distribution of Octets \eids and non-SNMPv3-conforming \eids.}
    \label{fig:hamming-weight}
\end{figure}

To better understand the randomness of Octets and
Non-SNMPv3-conforming, which is crucial to their ability to serve as
fingerprinting identifiers, we analyze their Hamming weight
distribution.
\Cref{fig:hamming-weight} shows the relative Hamming weight distribution for both formats.
The Hamming weight can be used as an indicator of randomness.
Thus, the expectation of a randomly generated number would have half of its bits
set to `1' and the other half set to `0'.
The number of ones, \ie the Hamming weight, for a large set of randomly generated numbers is therefore distributed according to the normal distribution $\mathcal{N}$ with a mean around half the length of the bit string.
To meaningfully compare Hamming weights of variable-length bit
strings,
we choose to display the relative Hamming weight, \ie the fraction of bits set
to `1'.
As \Cref{fig:hamming-weight} shows, the relative Hamming weight of the Octets format is centered around the mean of 0.5, indicating a mostly random source behind the generation of these \eids.
Non-SNMPv3-conforming \eids on the other hand seem to be distributed on not completely random input,
as the relative Hamming weight distribution has a positive skew, \ie more \eids with
this format have fewer than expected bits set to `1'.
For this reason, we apply a series of filters, which increase the confidence in the uniqueness of SNMP \eids.
Note that randomly generated \eids can still be persistent (\ie they remain the same for every query) and in fact we enforce \eid consistency in our filtering pipeline.

\subsection{Engine Time and Engine Boots}

In addition to the \eid we also use the \etime and \eboots SNMPv3
response 
fields for alias resolution.
By subtracting the \etime from the exact packet receive time, we can derive the \textit{\lastreboot} for each responsive IP.
The tuple of (\lastreboot, \eboots) serves as an additional strong identifier (cf. \Cref{sec:appendix:uniqueness}) in our alias resolution technique (cf. \Cref{sec:alias}), while also being useful for fingerprinting purposes (cf. \Cref{sec:fingerprinting}).

\begin{figure}[!bpt]
    \centering
    \includegraphics[width=\linewidth]{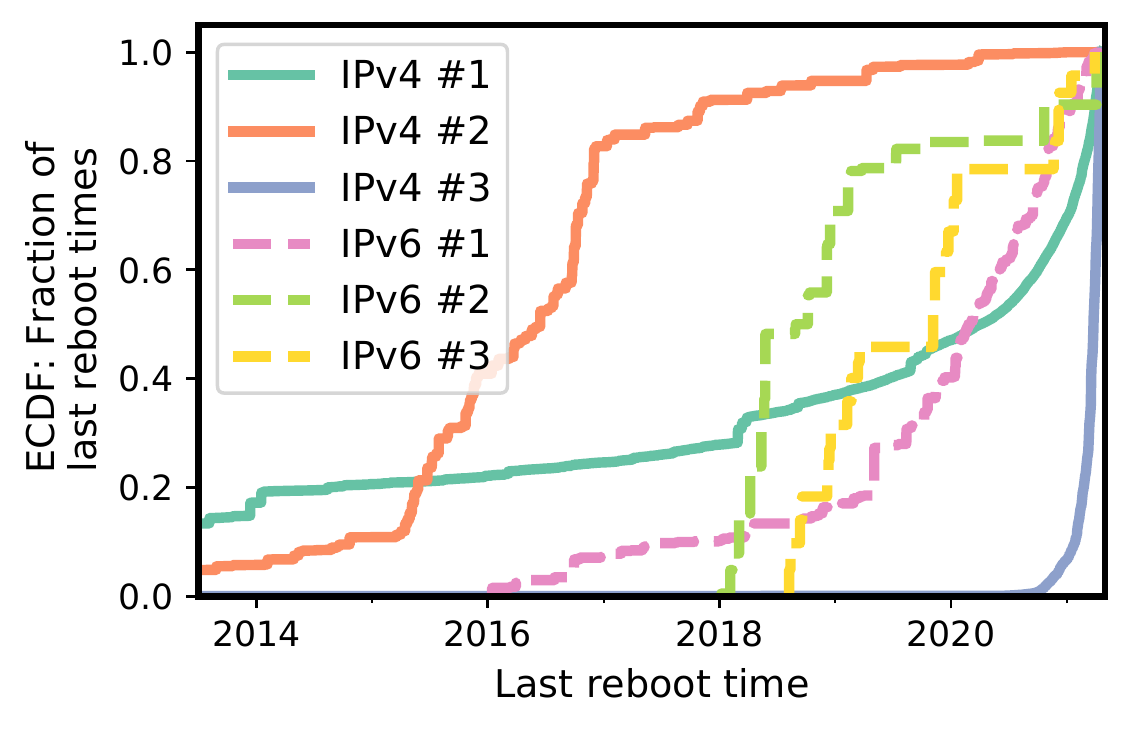}
    \vspace{-2em}
    \caption{Distribution of the \lastreboot for the top 3 \eids for IPv4 and IPv6.}
    \label{fig:enginetime-constant-engineid}
    \vspace{-1em}
\end{figure}

\Cref{fig:enginetime-constant-engineid} shows the distribution of the \lastreboot of the top three \eids for IPv4 and IPv6.
If those would belong to the same device, then we would expect to see them centered around the same time, \ie a single device with a unique \eid should have the same \lastreboot value.
We see, however, that five of the six most popular \eids have \lastreboot values spanning multiple years.
One prominent reason for this \eid reuse are software bugs in routers, as is highlighted by our \#1 IPv4 \eid which we find on more than 181k IPs.
This artifact can be traced back to a bug which was acknowledged by the vendor~\cite{cisco-bug-engineid}, resulting in a constant MAC-based \eid.
These examples underline the importance of using the tuple of
(\lastreboot, \eboots) as a second identifier in combination with the \eid.

\subsection{Filtering Responses}
\label{sec:datasets:filtering}

We perform multiple filtering steps for all SNMPv3
responses:

\noindent {\bf Missing \eids.} %
First, we remove responses with a missing \eid.
        This is mostly due to non-SNMP-compliant responses.
        With this filter we remove about 5k IPv4 and 15 IPv6 responses.

\noindent{\bf Inconsistent \eids.} %
In this step we merge the first and second scans for IPv4 and IPv6, respectively.
        Due to inconsistent answering behavior, likely due to devices
changing IP addresses in the interim time between our scans, we have an overlap of 30.2M IPv4 addresses out of the 31.8M and 31.5M for the first and second scan respectively.
        We remove an additional 1.4M responses which show inconsistent \eid values for the scans.
        In IPv6, 172k out of the 182k and 180k responsive IPs are overlapping, and we remove 557 instances of inconsistent \eid values.

\noindent{\bf Too short \eids.} %
As we rely on the uniqueness of \eids, we
filter responses with overly short \eids.
        We use a threshold of four bytes, in order to keep IPv4-based \eids in the data set.
        We ensure that IPv4 \eids provide enough uniqueness in the
        following steps.
        In this step we remove about 5\% for each protocol, \ie 1.5M for IPv4 and 10k for IPv6.

\noindent{\bf Promiscuous \eids.} %
        We leverage the enterprise ID, which is part of the \eid field and contains vendor information, to check for promiscuous \eid values.
        Specifically, we check whether the same \eid value is present across multiple vendors.
        If this is the case we label the \eid as promiscuous and as a result we remove 96k IPv4 and 555 IPv6 responses.

\noindent{\bf Unroutable IPv4 \eids.} %
        In this filtering step we check whether IPv4-address-based \eids actually contain routable IPv4 addresses.
        Non-routable addresses (\eg reserved, multicast, private addresses) are not guaranteed to be unique, and we therefore remove 68k IPv4 responses and and 7.8k IPv6 responses.

\noindent{\bf Unregistered MAC \eids.} %
        For all MAC-based \eids we map the contained MAC addresses to
      get the associated OUI~\cite{macoui}.
        We remove 113k and 1.4k MACs without matching OUIs for IP4 and IPv6, respectively.

\noindent{\bf Zero \etime or \eboots.} %
        We remove 834k IPv4 and 9.4k IPv6 entries with zero or empty \etime or \eboots values.

\noindent{\bf \Etime in the future.} %
        Next, we compare the \etime value with the packet receive time.
        As we rely on the \etime value reflecting a real time value, we remove entries where the \etime is in the future.
        In this step we remove 23k and 18 IPs for IPv4 and IPv6, respectively.

\begin{figure}[!bpt]
    \centering
    \includegraphics[width=\linewidth]{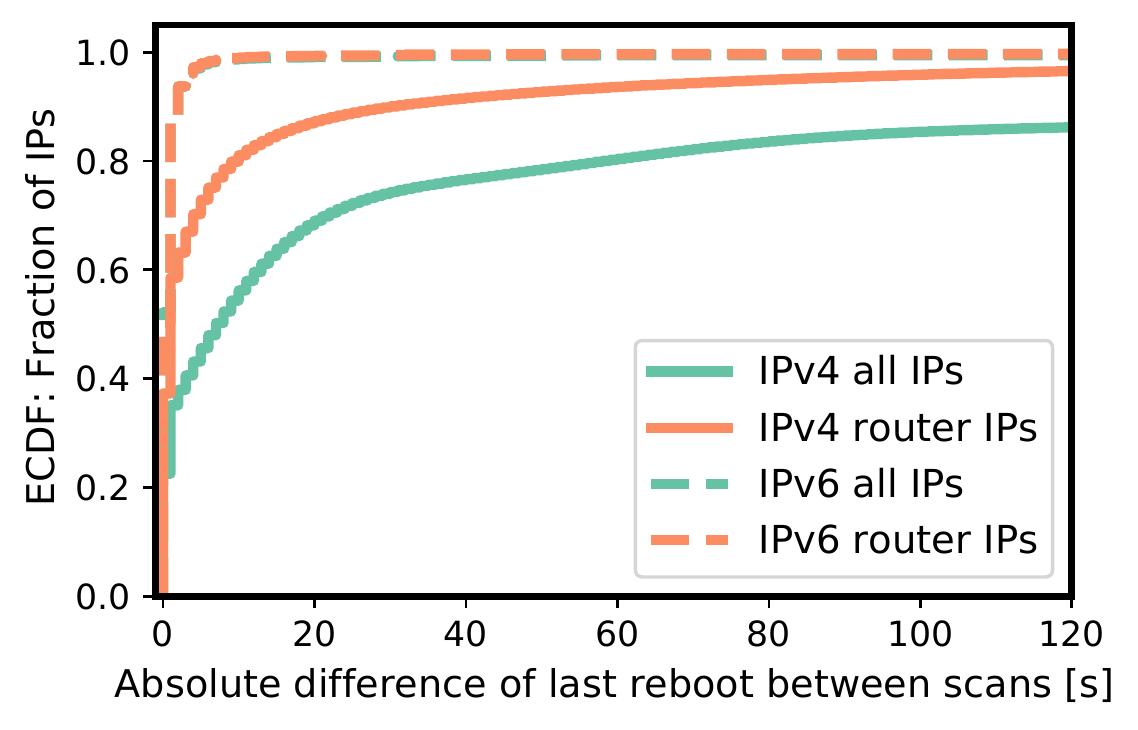}
    \vspace{-2em}
  \caption{Distribution of the \lastreboot difference between both scans for all IPv4, IPv6, and router IPv4, IPv6 addresses.}
    \label{fig:last-reboot-time}
\end{figure}

\noindent{\bf Inconsistent \eboots.} %
        In this step we compare the \eboots values for both scans.
        If they differ (\eg because of a reboot) we can not rely on the reset \etime value and we therefore remove 3.8M IPv4 entries and 802 IPv6 entries.
\noindent{\bf Inconsistent \lastreboot.} %
        We compare the derived \lastreboot value from both scans to check for consistency.
        As timekeeping is prone to clock skew \cite{IPv4-6-Siblings,zander2008improved,murdoch2006hot,kohno2005remote} and suddenly running clock synchronization daemons, we first analyze the difference of the last reboot time as shown in \Cref{fig:last-reboot-time}.
        As can be seen, the \lastreboot values in IPv6 overlap very consistently, while they are more spread out in IPv4.
        We selectively also depict the distribution for router IP addresses (cf. \Cref{sec:datasets:subsub:router}), which shows more consistent \lastreboot values.
        We choose a threshold of 10 seconds between scans, at the ``knee'' of the IPv4 router IPs distribution.
        With this last filter we remove 9.8M IPv4 addresses and 1.7k IPv6 addresses.

After this rigorous filtering pipeline we continue our analysis with the remaining 12.5M IPv4 addresses and 140k IPv6 addresses.
Although this is a significant decrease from the initial set of
SNMPv3-responsive IPs---especially for the more than 30M initial IPv4
responses---we err on the side of precision by applying this conservative filtering approach.

\section{Alias Resolution}\label{sec:alias}

Against the filtered IPv4 and IPv6 dataset, we run an alias resolution algorithm---first for IPv4 and IPv6 separately, and then on the combined set---to identify IP addresses belonging to the same SNMP device.
We try variations of our technique (cf. \Cref{sec:appendix:aliasresolution}) and choose an approach which mimics similar thresholds as our filtering pipeline described in the previous section.
We group all IP addresses together if they contain the same \eid, the
same \eboots, and a very similar \lastreboot for both scans.
In our filtering pipeline we select a \lastreboot threshold of 10 seconds.
To account for the fact that groups of IP addresses might deviate 10 seconds each, we map the \lastreboot time into 20 second bins.

\subsection{IPv4 and IPv6}

By grouping based on these six fields (\eid, \eboots, and \lastreboot, for both scans respectively) we create alias sets.
This results in 4.7M alias sets for IPv4, of which 824k are
non-singletons (\ie they contain more than one address).
As a result, more than 8.7M of the 12.5M (70\%) input IPs are grouped into non-singleton alias sets.
Each alias set contains on average 10.6 IP addresses.

For IPv6, we use the same technique and end up with 59k alias sets of
which 26k are non-singleton sets.
These non-singleton IPv6 alias sets contain more than 106k of the initial 140k IPv6 addresses, a coverage of more than 75\%.
Due to the lower number of responsive IPv6 addresses in our measurements, the average number of 4.2 addresses per IPv6 non-singleton alias set is smaller than in IPv4.

Finally, we also resolve dual-stack aliases (\ie devices with IPv4 and IPv6 addresses) by applying the same alias resolution technique on the joined IPv4 and IPv6 alias sets.
After this final alias resolution step we have 4.6M IPv4-only alias sets (\ie alias sets containing only IPv4 addresses), 27.7k IPv6-only alias sets, and 31.2k dual-stack alias sets.
Of those 796k, 11.3k, and 31.2k are non-singleton alias sets.
These non-singleton alias sets contain 7.4M IPv4-only addresses (9.3 addresses per set), 49.5k IPv6-only addresses (4.4 per set), and 1.4M dual-stacked addresses (45.4 addresses).
We find that especially the high number of average addresses for dual-stack alias sets is an impressive confirmation that our technique is able to reliably identify enterprise routers with many interfaces.

\begin{figure}[t]
    \centering
    \includegraphics[width=\linewidth]{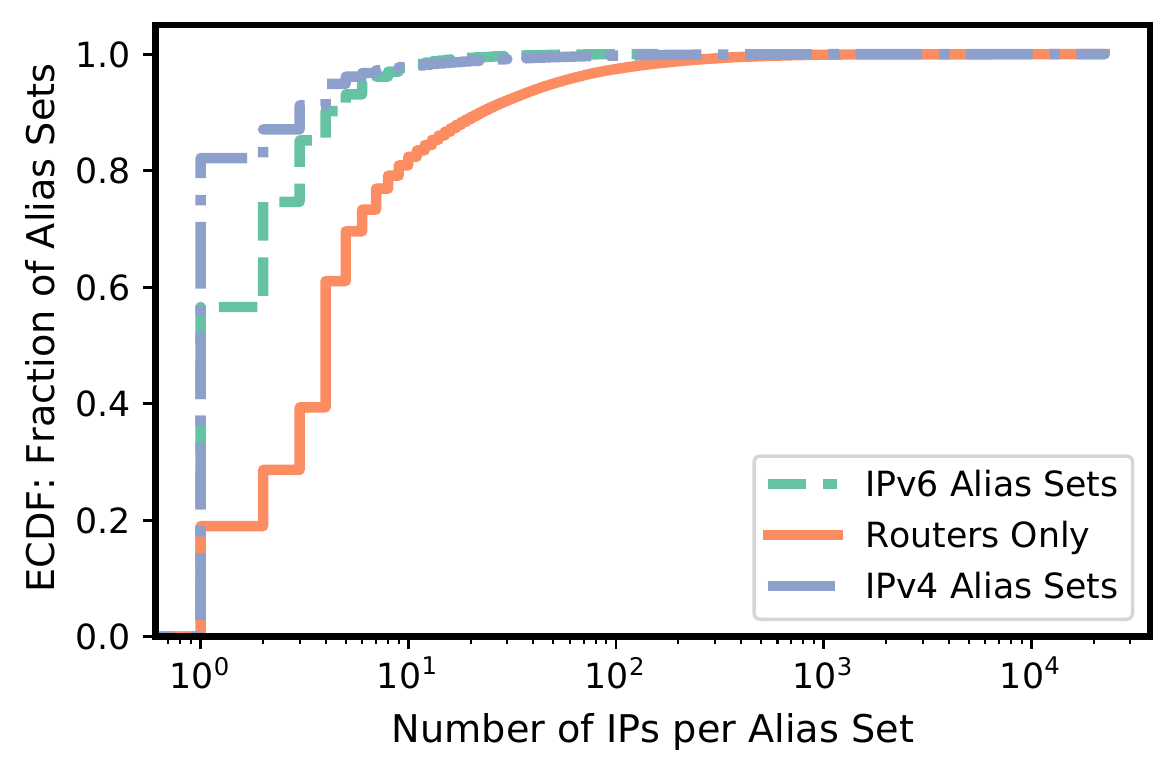}
    \caption{Distribution of the number of IP addresses per alias set for IPv4, IPv6, and router addresses.}
    \label{fig:ips-per-alias-set}
\end{figure}

In \Cref{fig:ips-per-alias-set} we show the distribution of IP addresses per alias set for IPv4, IPv6, and router IPs as identified by at least one router being part of a well-known router dataset.
In contrast to overall IPv4 and IPv6 alias sets, we find that router alias sets contain many more addresses.
This is an indicator that SNMPv3 is widely used on routers with many IP addresses and interfaces.

\subsection{Comparison with %
Router Names}

Next we compare our identified alias sets with the CAIDA Router Names dataset.
This dataset is built using the technique from Luckie \etal~\cite{regex19} by getting the reverse DNS name for IP addresses and then using regular expressions to group routers together.
Specifically, we obtain the per-domain suffix regular expressions (``regexes'') created by CAIDA and derived from their most recent March 2021 ITDK topology.
These regexes extract the hostname from a complete PTR record to identify the router; multiple interfaces with a common hostname are then assumed to be aliases of the same router.
Conservatively, we only use regexes where their algorithm produced a positive predictive value of 0.8 or higher.

We apply each suffix's regex to the IPv4 and IPv6 interface PTR records available in the March 2021 CAIDA ITDK.
Note that not all interface IP addresses have PTR records and we necessarily exclude these.
We then coalesce those names into routers that have both IPv4 and IPv6 interfaces with PTR records where the hostnames match (the interface names need not match).
In total, using these regexes, we obtain 12.4k dual-stack non-singleton alias sets containing 63.8k IP addresses, \ie 5.2 addresses per alias set.
This dataset is overall significantly smaller compared to our 838k non-singleton alias sets.
Even when looking at dual-stack non-singleton alias sets we identify more than 2.5x (31.2k).
We compare the content of the alias sets identified by both approaches and find only 9 exactly matching alias sets.
When looking for partial matches we identify 5.9k partially overlapping alias sets, \ie at least one IP address from CAIDA's Router Names dataset is in one of the SNMPv3 alias sets.
This finding shows that the SNMPv3 approach partially covers about half of all Router Names alias sets.
In summary, we conclude that both alias sets are complementary, as they contain only partially overlapping addresses.
This is likely due to the different used techniques, \eg some routers may not have useful rDNS entries, while others might not respond to SNMPv3.
Overall, however, the SNMPv3 technique is able to identify significantly more alias sets.

\subsection{Comparison with ITDK}

Next we compare our alias set results with CAIDA's ITDK, namely the March 2021 MIDAR dataset \cite{keys13midar} for IPv4 and the Speedtrap dataset \cite{Speedtrap} for IPv6.
Those datasets leverage the presence of a monotonically increasing IP ID value to identify aliases.
MIDAR identifies 8.4M IPv4 alias sets of which the overwhelming majority are singletons.
There are 94k non-singleton sets containing about 363k IP addresses, \ie 3.9 IPs per alias set.
Speedtrap groups IPv6 addresses into 525k alias sets---again, the majority are singletons, with only 5.3k alias sets with more than one address containing 13.6k addresses.
Our identified sets find 222k and 4.3k perfectly overlapping alias sets in MIDAR and Speedtrap, respectively.
More than 95\% of those overlaps are singleton sets.
Finally, we identify partial overlaps for 1.1M MIDAR and 533k Speedtrap alias sets, with the vast majority being singletons.
To summarize, MIDAR and Speedtrap also provide complementary views of aliased routes, likely due to different support of the used techniques.
Overall, we find almost a magnitude more non-singleton alias sets compared to both sets.

\subsection{SNMPv3 Coverage}

\begin{figure}[!bpt]
	\centering
	\includegraphics[width=\linewidth]{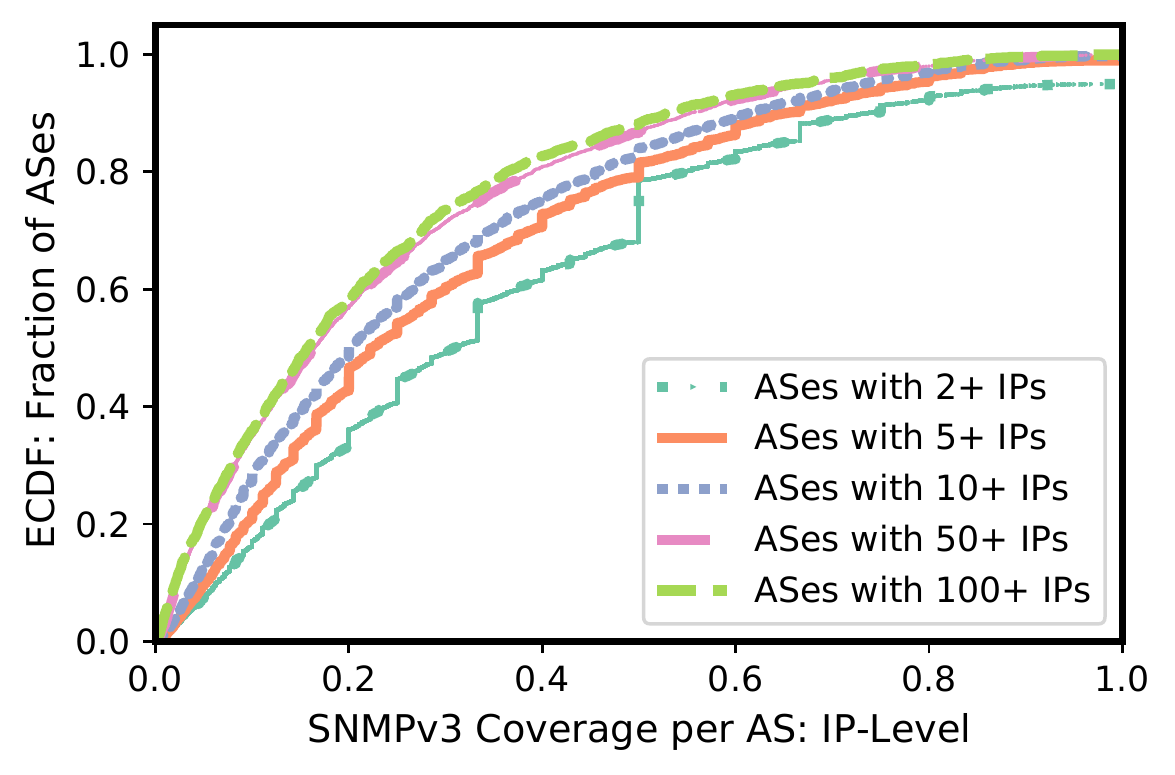}
    \caption{Coverage of responsive SNMPv3 router IPv4 addresses per AS.}
	\label{fig:coverage-as-ip-level}
\end{figure}

To assess how many IPv4 addresses per AS we can de-alias, we define {\it
coverage} as the ratio of responsive SNMPv3 router IPv4 addresses compared to the total number of IPv4
addresses per AS within the IPv4 union router dataset containing 3.1M addresses (cf. \Cref{table:router-datasets}).
Overall, we find that 16\% of the IPv4 router addresses respond to SNMPv3 probes.
In Figure~\ref{fig:coverage-as-ip-level} we plot the empirical CDF of SNMPv3 coverage
for ASes with thresholds of at least 2, 5, 10, 50, and 100 IPs in our dataset.
The number of ASes for each threshold are 11.8k, 9.1k, 7.9k, 2.9k, and 1.8k, respectively.
The main observation in Figure~\ref{fig:coverage-as-ip-level} is that SNMPv3 coverage is
slightly better for ASes with fewer IPs than those with higher ones. The coverage
also deviates substantially across different networks. 
Regardless of the threshold, the coverage is less than 10\% for about a quarter of the networks, and is
more than 80\% for top 10\% of the covered networks.
Recall, MIDAR contains around 94k non-singleton sets that consist of 362k router IPv4 addresses.
28.4\% of those addresses also respond to SNMPv3.
Further, more than 461k router addresses respond to SNMPv3 with only 22\% overlapping with MIDAR's non-singleton sets.
Overall, when combining both techniques one can potentially increase the coverage of de-aliased router IPv4 addresses from 11.7\% (MIDAR only) or 14.8\% (SNMPv3 only) up to 23\%.

\section{Router Fingerprinting}
\label{sec:fingerprinting}

\begin{figure}[!bpt]
    \centering
    \includegraphics[width=\linewidth]{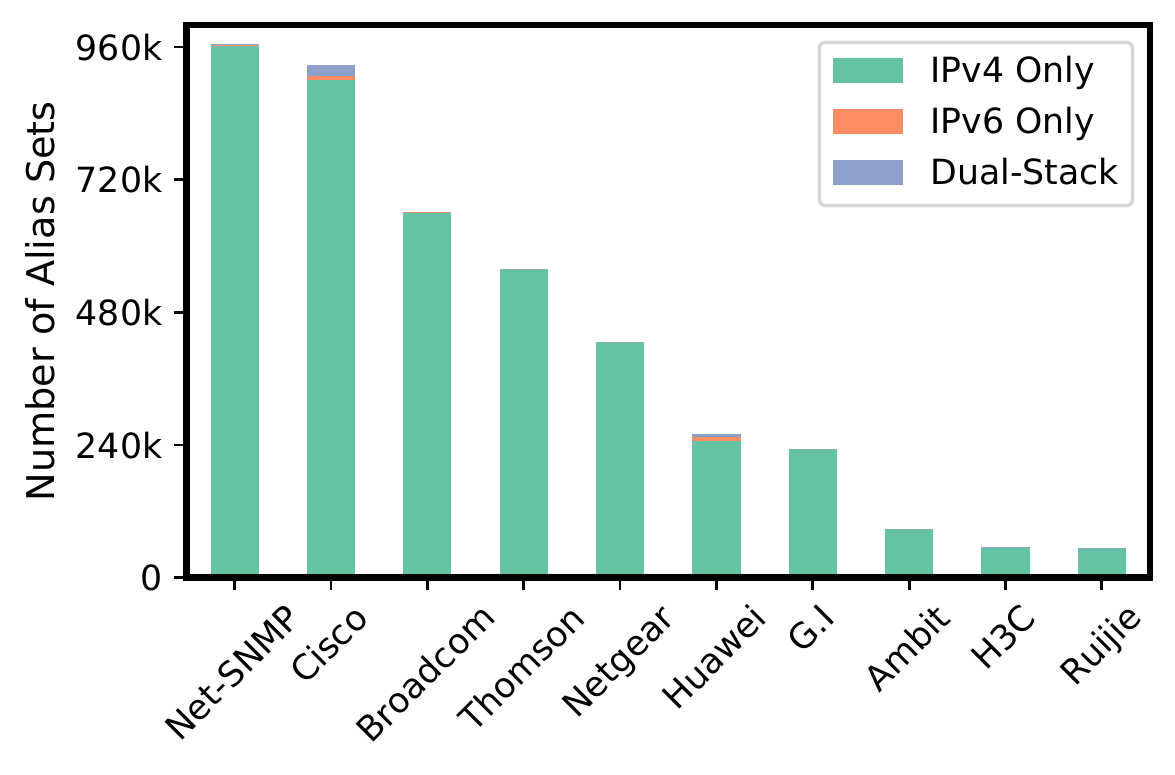}
    \caption{Vendor popularity.}
    \label{fig:fingerprinting-baseline-results} 
\end{figure}

Our technique offers the unique opportunity to identify the vendor of millions
of devices, including routers. These fresh insights allow us to estimate the market share of
infrastructure vendors at an unprecedented scale with our lightweight and accurate method. %

\subsection{Baseline Results}

In Figure~\ref{fig:fingerprinting-baseline-results} we report the popularity of
major network equipment vendors as unveiled with our method. In total, we are
able to de-alias 4,617,690 devices (aka alias sets). As illustrated in the
figure, the majority of these devices use exclusively IPv4.  The
largest fraction of the devices are UNIX-based (Net-SNMP).  Major network
equipment vendors are also in the top 10 list, including Cisco (more than 900k
devices), Broadcom (580k devices), Huawei (220k devices), and H3C (5k devices).
In the top 10 vendors there are also home and office network equipment vendors
such as Thomson (580k devices) and Netgear (420k devices). Thus, our technique
provides a view into many popular edge devices.  Unfortunately, attackers will
also have this view when they send unauthorized and unsolicited requests and can
exploit known vulnerabilities. The count of devices belonging to a given vendor
drops drastically past the top 10 vendors.  Indeed, the top 10 vendors are
responsible for more than 80\% of the devices we identify with our technique.

\begin{figure}[!bpt]
    \centering
    \includegraphics[width=\linewidth]{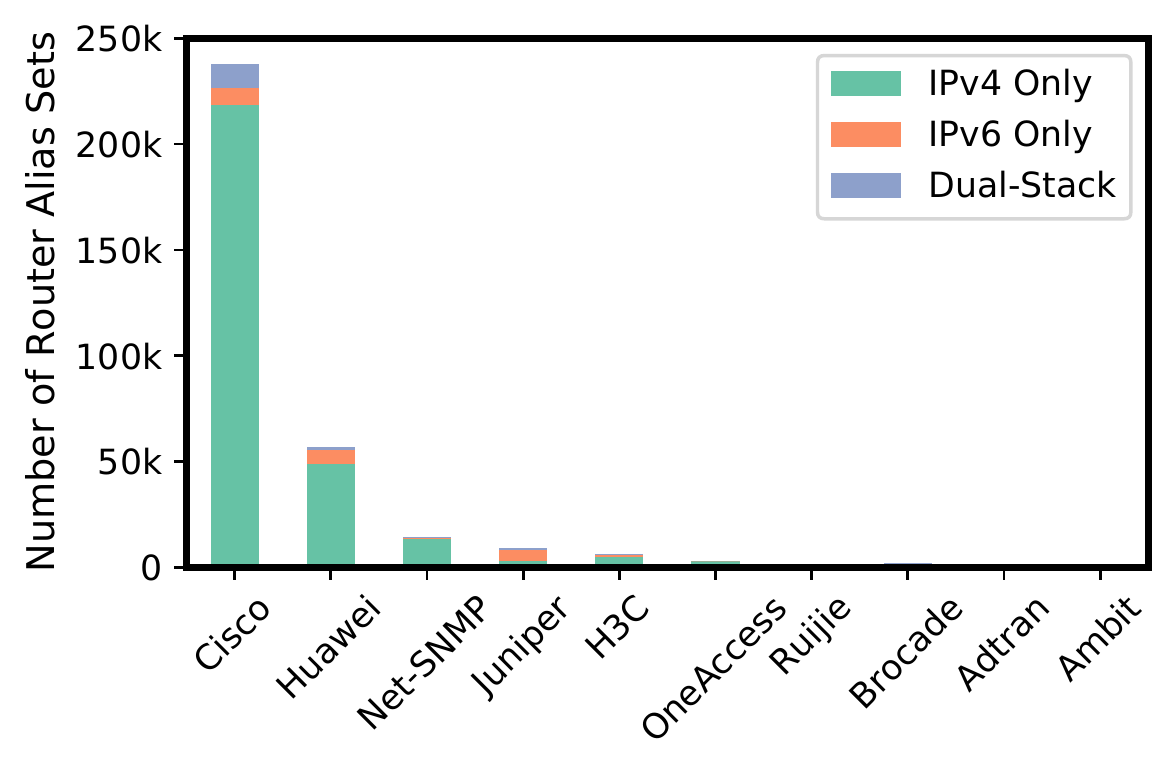}
    \caption{Router vendor popularity.}
    \label{fig:fingerprinting-baseline-results-routers}
\end{figure}

Next, we study the vendor popularity for the routers we identified with our
method. To compile the set of routers, we consider all the alias sets as before,
but we also require that the IPs in these alias sets are present in the most
recent ITDK and RIPE Atlas datasets, from March 2021 and April 2021,
respectively. In total, we identify 346,951 routers. The large majority are
IPv4-only (307,404), while there were 24,641 IPv6-only routers and 14,906
dual-stack routers. These numbers show that the fraction of IPv6
only and dual-stack routers is significantly higher for routers as compared to the
overall set of devices. This is also visible in
Figure~\ref{fig:fingerprinting-baseline-results-routers}, where we show the
popularity of router vendors by protocol. In this figure it is also clear that
when we consider routers there is higher vendor consolidation compared to the
overall devices. Indeed, the four major vendors, namely Cisco, Huawei, Juniper,
and H3C, are responsible for more than 95\% of routers we identified with our
method. Cisco is the most popular with around 240k routers followed by Huawei
with 52k routers.

\subsection{Validation}
\label{sec:fingerprinting:validation}

\subsubsection{Lab testing.} To better understand the default behavior
and configuration of Cisco devices running SNMPv3, we setup Cisco
routers running IOS 15.2(4)S7 (released in 2015) and IOS XR 6.0.1 
in a controlled
environment.  We use Net-SNMP to issue queries from a Linux machine
directly connected to the router and passively monitor the network
traffic via a packet capture.  By default, the Cisco router does not run SNMP
and answers neither v2 nor v3 queries.  We enable SNMP via a single
line of configuration \texttt{snmp-server community pass123 RO} which sets
the SNMP read community string to a private value.  We then confirm
that the router answers SNMPv2 queries using this private community
string by querying for the \texttt{sysDescr} MIB value and receiving the
response.

Next, we issue an SNMPv3 query, again for the \texttt{sysDescr} MIB object,
using the username \texttt{noAuthUser} and the security level
\texttt{noAuthNoPriv}, \ie the same unauthenticated query we issue in 
our Internet-wide measurements.  While the query is rejected with a
``unknown user name'' error as expected, for both versions of Cisco
IOS, the response packet includes a
Cisco OUI MAC address within the \eid field.  The router has
multiple interfaces, each with different MAC and assigned
IP addresses.  Regardless of the IP address queried, the router 
returns the same MAC address in the \eid response.  This MAC
address corresponds to the ``first'' interface as reported by the
routers via the command-line management interface. 
This MAC in the \eid was not the 
numerically smallest MAC address among
all the interfaces,
which contradicts the guidance in the SNMPv3 behavior
specification
\cite{IETF-RFC3411}.

Of note is that the Cisco router responds to these SNMPv3
queries without any specific v3 configuration.  Simply configuring
a read-only community string, which is only pertinent to SNMPv2, 
seemingly enables SNMPv3 on the router.  Thus, operators may 
inadvertently be enabling SNMPv3 on their devices in the process
of setting up SNMPv2.
We notified Cisco of our findings and they directed us to several
existing bugs and bugfixes which we discuss in~\Cref{sec:discussion}.

We replicate the same experiment with Juniper Junos (version 17.3, 2017
release). We notice that the behavior is similar to Cisco, \ie operation of
SNMPv2 enables SNMPv3. However, Juniper requires to explicitly enable services
on a given interface, which may result in less visible Juniper routers with our
SNMPv3 scan.

\subsubsection{Operators Survey.} To validate our results, we contacted network operators. In our request for
comment, we shared with them the alias sets, \ie the set of IPs for each
router and the router vendor as we identified it with our technique. Six of them replied to
our request. The network operators confirmed that we correctly
de-alias the router and identified the router vendor in all responses. 
We also notice that identified ``net-snmp'' and ``unknown vendor'' devices, typically correspond to
network appliances and programmable network devices, respectively.  
However, some of the operators pointed out potential limitations of our method. Indeed, we were unable to uncover some router interfaces with our SNMPv3 scans, as those ACL interfaces drop packets sent to well-known ports.

\begin{figure}[t]
    \centering
    \includegraphics[width=\linewidth]{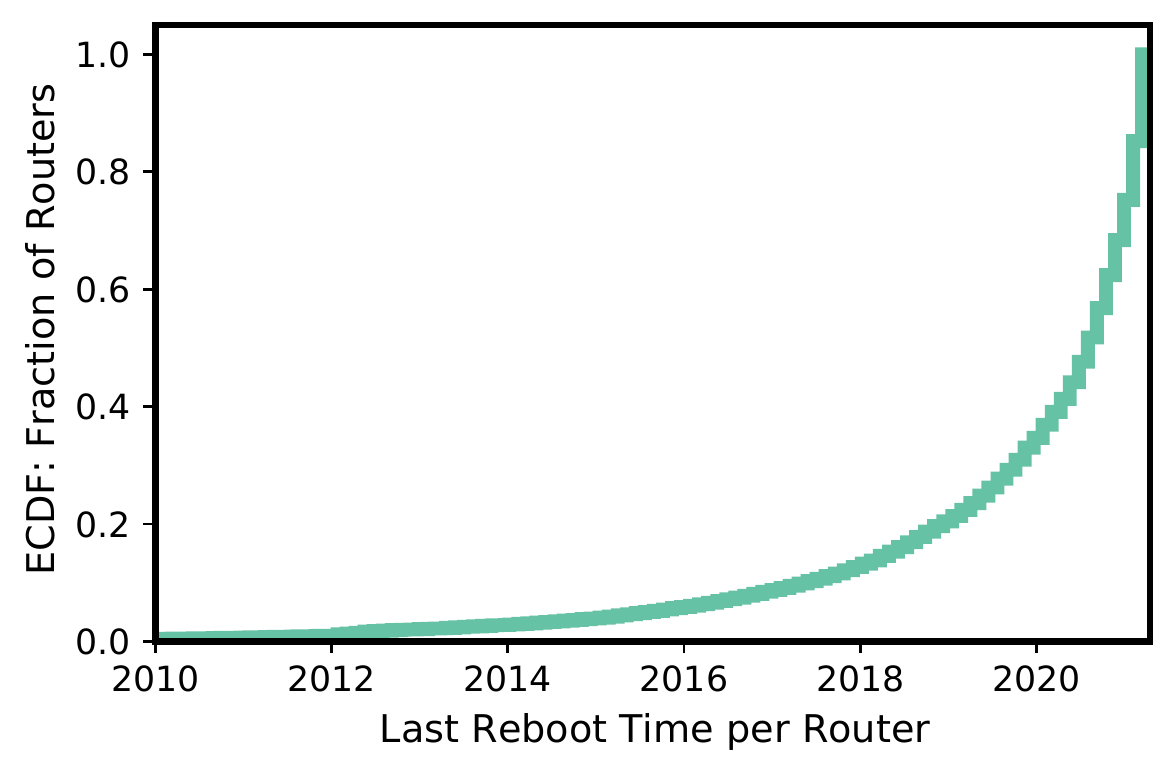}
    \caption{Distribution of time since the last reboot.}
    \label{fig:router-time-since-last-reboot}
\end{figure}

\subsubsection{Comparison with Nmap.}

Nmap~\cite{Nmap} is a popular tool for operating system and
device fingerprinting. It runs up to 16 TCP, UDP, and ICMP tests in order to
generate a signature for a target, and then attempts to match it against its
database. Nmap requires at least one open and one closed TCP port to achieve
accurate results. If Nmap is unable to run all the required tests, or find an
exact match to the resulting signature, it attempts to provide a best-guess of
the target OS. Unsurprisingly, Nmap's approach generates a significant amount of
traffic and is not suitable for a large-scale measurement. As such, we decide
to test it only on a small subset of all SNMPv3 responsive routers. We randomly
pick a single IPv4 address from each router. We target 26.4k router IPs in total
and compare the resulting fingerprints from Nmap (version 7.60) with the one
obtained via SNMPv3. 

For 22.2k routers, Nmap was unable to report any results. This is likely due to
the fact that none of those routers are running any common TCP service (ftp,
ssh, telnet, etc.) which is required for Nmap to work properly. Note that by
default, Nmap will attempt to find an open TCP port by scanning only the top 10
services in its database. We acknowledge that Nmap may fingerprint those routers
when using more aggressive options (such as full TCP port scan) but we opted not
to do so due to the excessive load it would generate. Further, the Nmap fingerprint did not agree
with our SNMPv3 fingerprint for 1.3k routers. In all of those instances, Nmap
attempted to guess the operating system rather than providing an exact match
from its database indicating that it was unable to complete some of its tests.
Finally, 2.9k Nmap results match the fingerprint obtained with SNMPv3; Nmap
was able to provide additional information such as OS version for the majority
of those routers. Recall that our method is not using any statistical inference
as our vendor identification is based on either the MAC address or Enterprise ID---that are typically unique per vendor ~\cite{EnterpriseNumber}---both of which
can be obtained from the SNMPv3 \eid data. We acknowledge that Nmap's  thorough
tests and large database can fingerprint devices beyond the vendor level.
Nevertheless, contrary to Nmap the SNMPv3 technique allows for Internet-scale fingerprinting by sending only a single probe packet per target address.

\subsection{Time Since Last Reboot}

Accurate router fingerprinting allows us to answer important questions about the
status of routers in the wild.
In Figure~\ref{fig:router-time-since-last-reboot} we
plot the CDF for the time since the last reboot for around 346k routers we
identified with our method. Less than 25\% of them had their last reboot 
more than a year before our first scanning campaign (ca. April 2021).
More than 50\% of the routers had a reboot since the beginning of the year of our
measurement (2021), and around 20\% during the last month. These results show
that, potentially, a large fraction of routers did not recently install security updates, for which a reboot is normally necessary.
We are currently launching more campaigns and we will continue monitoring the last
reboot time to provide more insights in the future.

\begin{figure}[t]
    \centering
    \includegraphics[width=\linewidth]{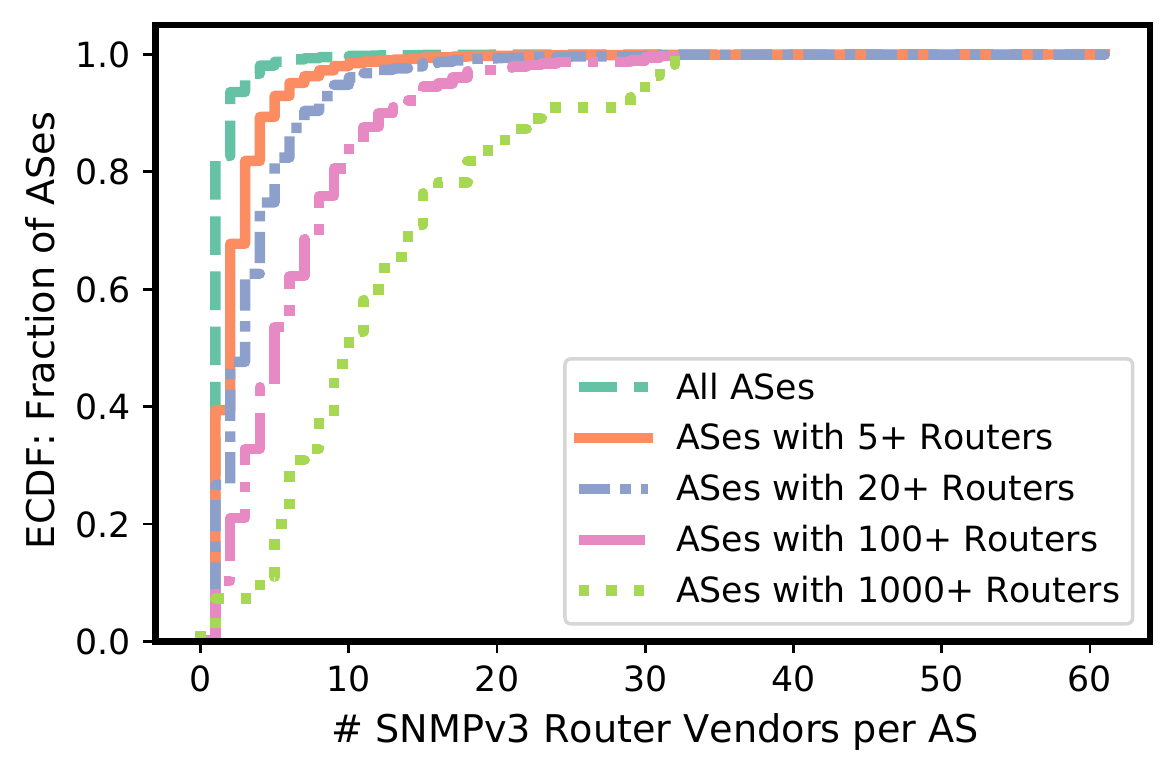}
\caption{Number of router vendors per AS.}
    \label{fig:router-vendors-per-AS}
\end{figure}

\subsection{Router Deployment Insights}

\subsubsection{Distribution of Routers per AS}
We consider router deployments in 22,787 networks. In 4,059 networks we identified at least 5 routers, in 1,557 networks more than 20 routers, in 381 networks more than 100 routers, and in 55 more than 1,000 routers using our SNMPv3 scans. Our analysis shows that the distribution of routers per AS is similar across different regions, see
Appendix~\ref{sec:dist-routers-AS-region}.

\subsubsection{Number of Vendors per AS}

In Figure~\ref{fig:router-vendors-per-AS} we report how many different router
vendors we can identify in a single AS. In 40\% of the
networks with more than 5 routers all of the routers are from the a single
vendor. In less than 10\% of the networks the number of router vendors exceeds
five. When we are focusing on larger networks, with more than 100 or 1,000
routers, then we observe that there are more routers vendors present. This is to
be expected as these networks run complex network operations and they can host
specialized routers and network equipment from different vendors.
In Section~\ref{sec:top10nets} we do a case study of the 10 largest networks by
number of routers and we show that although the number of vendors may be high,
the majority of the routers are from a couple of vendors. This is also true in
most of the networks with 5 routers or more in our study.

\begin{figure}[t]
    \centering
    \includegraphics[width=\linewidth]{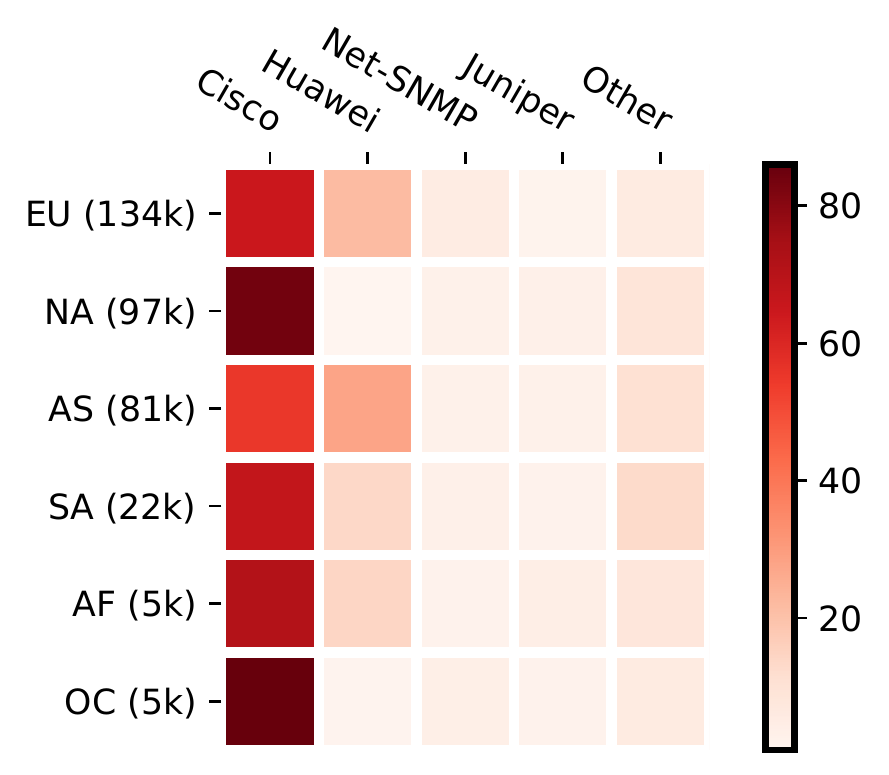}
    \caption{Router vendor popularity per continent with the total number of routers per region in parenthesis.}
    \label{fig:uniformity-regions}
\end{figure}

\subsubsection{Regional Vendor Popularity}

We then focus on the market share of different router vendors at different
regions. In Figure~\ref{fig:uniformity-regions} we present a heat map for the
popularity of each vendor in all the ASes of a region (continent). Cisco is the dominant
vendor across all regions. The second most popular is Huawei with about 27\%
market share in Asia, around 22\% in Europe, and close to 14\% in South America and Africa.
However, we could not find any Huawei router in North America and less than 1\% in
Oceania. The contributions of other router vendors is very low across regions.
We conclude that although the number of vendors per AS may be high
only a relatively small number of routers contribute to this diversity, as
there is a strong consolidation driven by two major router manufacturers, Cisco
and Huawei.

\subsubsection{Vendor Popularity in Large Networks}\label{sec:top10nets}

\begin{figure}[t]
    \centering
    \includegraphics[width=\linewidth]{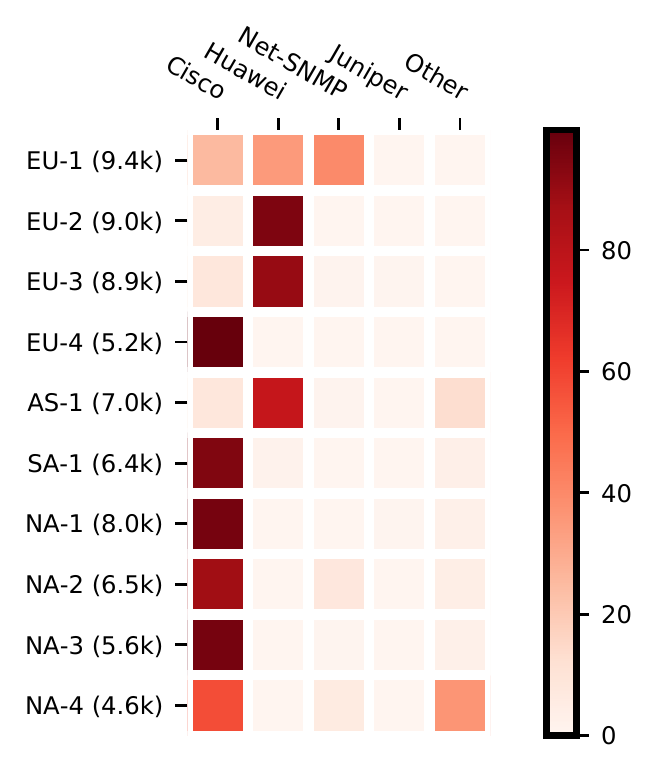}
    \caption{Router vendor popularity for top 10 networks by number of routers with number of routers in parenthesis.}
    \label{fig:uniformity-top100-networks}
\end{figure}

Finally, we perform a case study for the top 10 networks by number of routers. 
These are networks with at least 4.5k routers. They are distributed in different
continents, four in Europe, four in North America, and one in Asia and South
America. In Figure~\ref{fig:uniformity-regions}, we report the popularity of
each vendor and in parenthesis the number of routers per AS.
In 6 out of 10 of these networks Cisco is dominant. However, in the network from
Asia and in two of the networks in Europe, Huawei is dominant. Typically, the
large networks have only one dominant vendor, however, one of the networks in
the top 10 list has deployed both Cisco and Huawei routers as well as UNIX-based
router solutions. Again, although we see multiple router vendors present in all top 10
networks, the large majority of the routers (typically, more than 95\%) belongs to only one
or two vendors. Thus, large networks can be potentially vulnerable to
vulnerabilities of a single major vendor as their deployment is quite uniform.

\subsection{Vendor Dominance}

\begin{figure}[t]
	\centering
	\includegraphics[width=\linewidth]{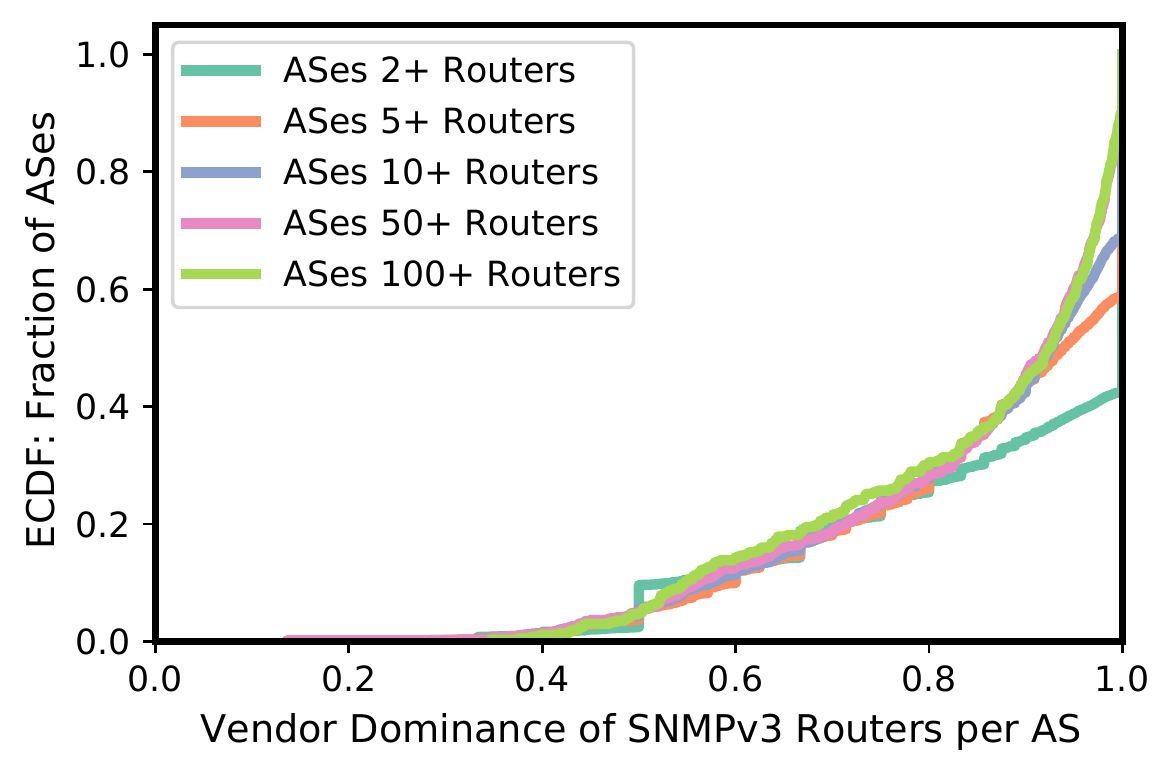}
    \vspace{-2em}
    \caption{Vendor dominance for routers found with SNMPv3 measurements per AS.}
	\label{fig:Vendor-dominance-as}
\end{figure}

\begin{figure}[t]
	\centering
	\includegraphics[width=\linewidth]{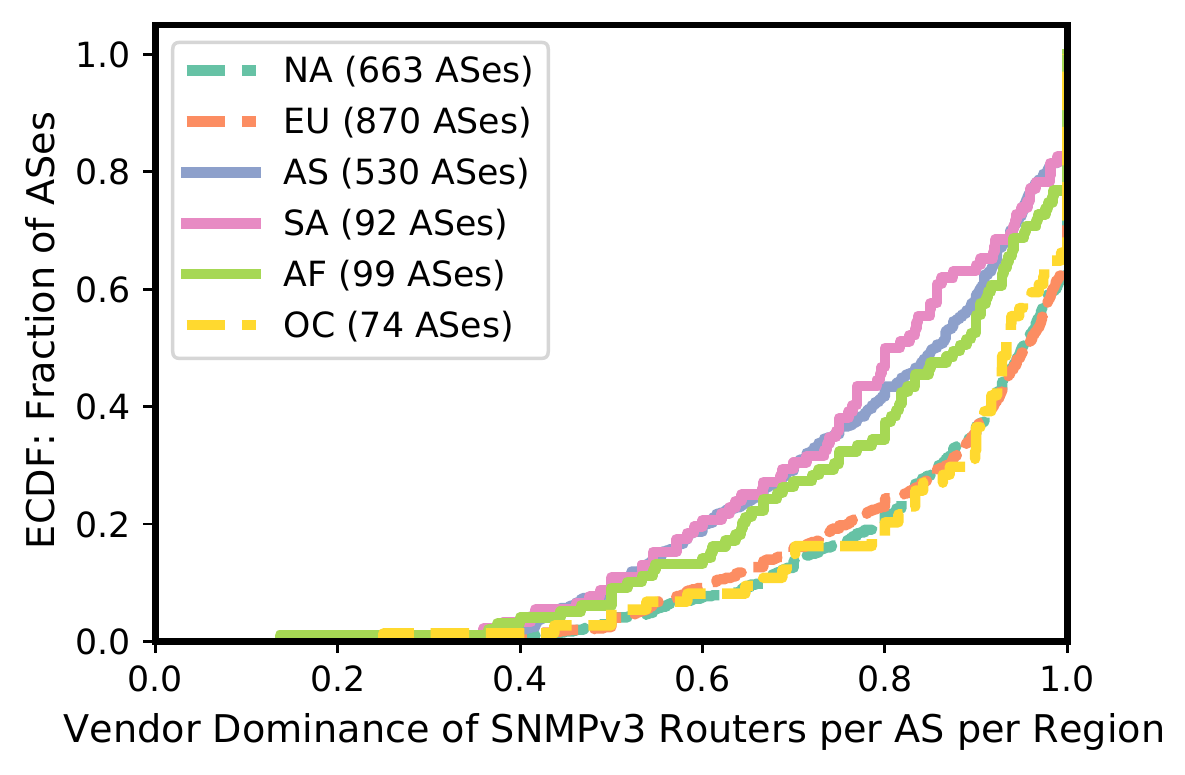}
    \vspace{-2em}
	\caption{Vendor dominance for routers found with SNMPv3 measurements per
    region for ASes with at least 10 routers.}
	\label{fig:Vendor-dominance-region}
\end{figure}

We also study the homogeneity of router vendors in each network.
In order to analyze this homogeneity we introduce a new metric, {\em vendor dominance}, \ie the fraction of routers belonging to the most dominant router vendor in that AS.
Thus, the closer the vendor dominance is to 1, the higher higher is the share of routers belonging to only a single vendor in this network.
This is a critical property
of a network, as vulnerabilities of the dominant vendor may be exploited and
affect a large fraction of the deployed routers.
In Figure~\ref{fig:Vendor-dominance-as} we show the
distribution of the vendor dominance across ASes.
We notice that the the values of vendor dominance are
high throughout many networks: more than 80\% of the networks have a vendor dominance of 0.7
or more. This shows that, typically, there is a single very popular vendor per network,
that can well be different from network to network.  Next, we turn our attention
to the regional characteristics of networks with at least 10 routers as measured
with SNMPv3, see Figure~\ref{fig:Vendor-dominance-region}. We notice that there
are two groups of regions: (i) South America, Asia, Africa, and (ii) Oceania,
North America, Europe. The vendor dominance of networks in the first group is typically
lower than in those of the second group. %

\section{Related Work}\label{sec:related}

Prior work has developed passive and active techniques
that leverage identifiers and implementation-specific differences to
fingerprint and de-alias devices at various granularities. This section details
these existing methods.

\subsection{Router Vendor Fingerprinting}

\noindent\textbf{Nmap:}
Nmap is an open-source network scanning and reconnaissance tool
that can perform operating system fingerprinting~\cite{Nmap}.  It
sends a series ICMP echo requests, UDP packets, and TCP
probes with different field values, flags, and options to fingerprint the
remote system.  By examining the responses, \eg length,
options, window size, sequence numbers, IP ID and TTL values, checksum,
and flags, Nmap finds the best matching implementation in its database
of operating system fingerprints.  The latest version of Nmap (7.91)
contains 5,679 fingerprints; of these, approximately 160 and 22
correspond to Cisco and Juniper routers respectively.
While Nmap is a powerful tool for TCP/IP fingerprinting, it requires
the remote host to listen and respond on an open TCP port.  Because
routers in the wild are secured and typically do not respond to
unsolicited TCP probes, Nmap is generally ineffective for router
fingerprinting.

\noindent\textbf{Scanning:}
A second popular technique for remotely inferring operating system and
vendor information is ``banner grabbing,'' whereby publicly available
services leak information.  For instance, the Cisco SSH server implementation returns
identifying information in its response string.  Internet-wide
scanning and banner grabbing are performed
regularly~\cite{ZMap,search-engine-CCS2015,Censys,LZR2021}.  Recent work
has sought to find banners and augment with active and analysis of
banners and augmentation with active
measurements~\cite{Classifying-Vendors} to perform large-scale network
vendor inference.  Unfortunately, as with Nmap's reliance on TCP,
banner analysis requires a publicly responsive service that returns
discriminating information.  Routers are frequently tightly secured
and unresponsive to banner queries.

\noindent\textbf{TTL:} 
Due to the relatively closed nature of routers, Vanaubel \etal
developed a fingerprinting technique based solely on TTL responses
\cite{TTL-fingerprinting}.  They send TCP, UDP, and ICMP probes 
toward the target, and show that the tuple of inferred
initial TTL (iTTL) values from the responses can
differentiate between some well known platforms, including Cisco and
Juniper.   Unfortunately, the universe of possible iTTL values is
small, and can lead to a large number of incorrect inferences, \eg
Huawei has the same iTTL signature as Cisco.  

\subsection{Router Alias Resolution}

\noindent\textbf{Identifiers:}
Techniques for discovering router aliases %
were
first developed by exploiting an implementation behavior where ICMP
port unreachable responses were generated with a source address of the
interface toward the destination, regardless of which interface IP was
originally probed~\cite{govindan2000heuristics}.  Other techniques
have included pre-specified
timestamps~\cite{sherry2010resolving} and graph
analysis~\cite{gunes2006analytical}.  More recently, Marder proposed
using path length estimation~\cite{marder2020apple} while Vermeulen \etal leverage
ICMP rate limiting~\cite{vermeulen2020alias}.  However, to date, 
state-of-the-art systems use IP ID based alias resolution.

\noindent\textbf{IP ID:} 
The IP identifier (IP ID) field is used for fragmentation and
reassembly.  Since IP ID is a mandatory field in IPv4, it is possible
to elicit an IP ID value from a router via a simple ICMP echo. %
This value is typically set in one of
three ways: zero (in conjunction with the ``don't fragment'' bit),
randomly, or sequentially for each new packet.  While this information
can convey some information about the router vendor, IP IDs can be used
for alias resolution---the process of determining if multiple IP
addresses correspond to different interfaces on the same router (\ie
are aliases).  In many implementations, the IP ID counter is shared
among different interfaces on the router~\cite{Rocketfuel}.  Current state-of-the-art
alias resolution techniques sample the IP ID value of candidate IPs
over time and perform a monotonic bounds test on IP ID sequences to
infer aliases~\cite{keys13midar}.  Similar techniques have been
developed for IPv6~\cite{Speedtrap}.  As the IP ID
is only 16 bits long, it can increment and wrap faster
than it can be sampled if the router has a large traffic rate.
As a result, IP ID techniques suffer from false positive and
false negative errors~\cite{regex19}.  Further, finding aliases based
on IP ID at Internet-scale remains challenging, and requires intensive
probing and computation.  In contrast, our technique elicits a strong,
unique, and persistent identifier that can discover aliases much more efficiently.

\subsection{Dual-stack Inference}

Comparatively less work has made progress on finding dual-stack
aliases, especially for routers.  Scheitle \etal use TCP timestamp
skew to identify IPv4/IPv6 siblings, but, as routers are generally
unresponsive to TCP probes, their work largely centers on
servers~\cite{IPv4-6-Siblings}.  Berger \etal developed a method to 
reveal dual-stack addresses of name servers ~\cite{berger2013internet}.
And while Czyz \etal examined
IPv6 security posture as compared to IPv4~\cite{Back-Door-IPv6}, 
they too had no method to resolve dual-stack router aliases.  To date,
the only viable technique for dual-stack router alias resolution
has been through DNS inference, in particular using Luckie's 
method for extract router hostnames~\cite{regex19} which we compare
against.

\section{Discussion}\label{sec:discussion}

\noindent {\bf Remediation.} In consultation with Cisco's product
security incident response team, we learned that they had multiple
existing bugs \cite{cisco00727,cisco74132} (the oldest one is from 2012) along with a reserved CVE \cite{cve} relating to SNMPv3 information leakage.
As a fix, Cisco added a new
configuration command that instructs the router to drop (and not
respond to) SNMP packets for an unknown user (specifically:
\texttt{snmp-server drop unknown-user}).  While this patch is welcome,
our findings underscore the fact that, despite this vulnerability
being known for nearly a decade, it persists in the wild.  Further,
operators must know of, and explicitly enable, this configuration
option in their equipment.  As such, we recommend that this
configuration become the default in the future.

\noindent {\bf Follow Best Current Security Practices When Running SNMPv3.}
To our surprise, millions of devices, including hundreds of thousands of
routers, respond to unauthorized and unsolicited SNMPv3 requests. Recall that
all of our requests were
launched by a server in a single data center, implying that these devices either
were not behind a firewall or the firewall was not properly configured. If the
network administrators had applied best current security practices, e.g.,
access control lists or
segregated out-of-band management, 
we would expect significantly fewer responses.
This would
have mitigated the potential vulnerabilities we identified at large.
Similarly surprising, as mentioned in
Section~\ref{sec:methodology}, we did not receive any
complaints
(although we provided contact information) when we scanned routers, in
striking contrast to scans performed in the past for servers. This can be
attributed to the fact that SNMPv3 scans are stealthy as they send only a single UDP
packet, but it also suggests that routers and other connected devices that
run SNMPv3 are not well monitored by network administrators for scanning
activity.  Operators may want to more rigorously monitor such external
management queries to their infrastructure.
\noindent {\bf Implicitly Enabling SNMPv3.}
In addition to the open network access policies previously identified,
a likely explanation for the large number of responsive SNMPv3 devices
is that some vendors and implementations automatically enable SNMPv3 when SNMPv2 is
enabled (see \Cref{sec:fingerprinting:validation}).  Instead, we 
recommend that implementations require explicit configuration to 
enable SNMPv3 to ensure that operators are consciously running SNMPv3.

\noindent {\bf Potential Vulnerabilities of Current SNMPv3 Implementation.}
There are more than 400 vulnerabilities related to SNMP~\cite{snmp-cve}. 
While many of these vulnerabilities are related to specific
implementations, our observations on SNMPv3's behavior exposes a more
fundamental fingerprinting weakness.  Whereas an unauthorized and
unsolicited request in SNMPv2 does not elicit a response, SNMPv3, by
design, does respond.  
Our study shows that
potentially millions of devices will respond to such SNMPv3 requests. 
Moreover, as SNMPv3 is UDP-based, it is trivial to spoof the source of
these requests, akin to other spoofed-source
attacks~\cite{spoofer-imc09,rossow2014amplification,sargent2017potential,gasser2017amplification}.
More concerningly, in some of our measurements,
a single request resulted in multiple (identical) responses.
For example, in the
the first IPv4 scan, more than 182k IPv4 addresses responded with more than one
request, 48 of which returned more than 1,000 responses within two to twenty
hours. One of them, sent back 48,500,523 response packets within two hours. 
Although the exact cause of this behavior is not
known, similar behavior has been reported for other handshaking
protocols~\cite{Hell-Handshake}. In the second IPv4 scan, we also saw different 
addresses return a large number of responses. Thus, SNMPv3 could be potentially
exploited for amplification attacks.
\noindent {\bf Towards SNMPv4.} Although SNMPv3
offers a stronger security model
than its predecessors, our study uncovered shortcomings in the
protocol that enable fingerprinting.
A fundamental design tension exists between 
maintaining the stateless operation of SNMP and the security mechanisms.
While proposals to utilize standards such as TLS with SNMP have 
been drafted to protect datagrams~\cite{rfc6353}, such approaches may present
other concerns, including the ability to perform certificate-based
fingerprinting.  A more immediate solution is to not use MAC addresses
as the \eid.  While the \eid needs to be persistent, re-purposing
MAC addresses has a long history in network security of enabling 
fingerprinting and
other attacks.
Further, 
researchers have shown that obtaining the persistent
\eid permits brute force SNMPv3 password recovery
attacks~\cite{SNMPv3-bruteforce-attacks}.  We thus encourage protocol
designers in the future to consider the weaknesses of a persistent
\eid, as well as discourage the use of MAC address-based \eid.

\section{Conclusion}\label{sec:conclusion}

In this paper, we show that the adoption of a secure network management
protocol, SNMPv3, surprisingly increases device fingerprinting capabilities.  By
design, devices that run SNMPv3 respond to unauthorized and unsolicited
authentication requests with a device unique identifier and other critical
status and configuration information.  We show that
SNMPv3 allows for light and accurate alias resolution, dual-stack
association, and fingerprinting with only a single request per
IP. With our technique we were able to de-alias
and fingerprint more than 4.6 million devices, including around 350k routers.
Our analysis provides fresh insights on the router vendor share, router deployment strategies of
network operators around the world, as well as router
uptime statistics and distribution of vendors in different regions. We hope that our
technique can be used for answering other network analytics questions in the
future, e.g., inferring NAT and load balancers in the wild. %
\subsection*{Acknowledgments}

We thank Cas D'Angelo for providing invaluable operational network validation.
We would also like to thank our shepherd Mattijs Jonker and the anonymous reviewers for
their valuable feedback.  This work was funded in part by the European Research
Council Starting Grant ResolutioNet (ERC-StG-679158), BMBF BIFOLD 01IS18025A and
01IS18037A, and the U.S.\ National Science Foundation (CNS-1855614).  Views and
conclusions are those of the authors and should not be interpreted as
representing the official policies or position of the ERC, BMBF, U.S.\
government or the NSF.

\label{page:end_of_main_body}

\bibliographystyle{ACM-Reference-Format}
\bibliography{paper}


\begin{thebibliography}{61}


\ifx \showCODEN    \undefined \def \showCODEN     #1{\unskip}     \fi
\ifx \showDOI      \undefined \def \showDOI       #1{#1}\fi
\ifx \showISBNx    \undefined \def \showISBNx     #1{\unskip}     \fi
\ifx \showISBNxiii \undefined \def \showISBNxiii  #1{\unskip}     \fi
\ifx \showISSN     \undefined \def \showISSN      #1{\unskip}     \fi
\ifx \showLCCN     \undefined \def \showLCCN      #1{\unskip}     \fi
\ifx \shownote     \undefined \def \shownote      #1{#1}          \fi
\ifx \showarticletitle \undefined \def \showarticletitle #1{#1}   \fi
\ifx \showURL      \undefined \def \showURL       {\relax}        \fi
\providecommand\bibfield[2]{#2}
\providecommand\bibinfo[2]{#2}
\providecommand\natexlab[1]{#1}
\providecommand\showeprint[2][]{arXiv:#2}

\bibitem[\protect\citeauthoryear{Authority}{Authority}{2021}]%
        {EnterpriseNumber}
\bibfield{author}{\bibinfo{person}{Internet Assigned~Numbers Authority}.}
  \bibinfo{year}{2021}\natexlab{}.
\newblock \bibinfo{title}{{PRIVATE ENTERPRISE NUMBERS}}.
\newblock
  \bibinfo{howpublished}{\url{https://www.iana.org/assignments/enterprise-numbers/enterprise-numbers}}.
\newblock


\bibitem[\protect\citeauthoryear{Berger, Weaver, Beverly, and Campbell}{Berger
  et~al\mbox{.}}{2013}]%
        {berger2013internet}
\bibfield{author}{\bibinfo{person}{A. Berger}, \bibinfo{person}{N. Weaver},
  \bibinfo{person}{R. Beverly}, {and} \bibinfo{person}{L. Campbell}.}
  \bibinfo{year}{2013}\natexlab{}.
\newblock \showarticletitle{{Internet nameserver IPv4 and IPv6 address
  relationships}}. In \bibinfo{booktitle}{\emph{ACM IMC}}.
\newblock


\bibitem[\protect\citeauthoryear{Beverly}{Beverly}{2002}]%
        {RTG}
\bibfield{author}{\bibinfo{person}{R. Beverly}.}
  \bibinfo{year}{2002}\natexlab{}.
\newblock \showarticletitle{{RTG: A Scalable SNMP Statistics Architecture for
  Service Providers}}. In \bibinfo{booktitle}{\emph{{USENIX} LISA}}.
\newblock


\bibitem[\protect\citeauthoryear{Beverly, Berger, Hyun, and k.~claffy}{Beverly
  et~al\mbox{.}}{2009}]%
        {spoofer-imc09}
\bibfield{author}{\bibinfo{person}{R. Beverly}, \bibinfo{person}{A. Berger},
  \bibinfo{person}{Y. Hyun}, {and} \bibinfo{person}{k. claffy}.}
  \bibinfo{year}{2009}\natexlab{}.
\newblock \showarticletitle{{Understanding the Efficacy of Deployed Internet
  Source Address Validation Filtering}}. In \bibinfo{booktitle}{\emph{ACM
  IMC}}.
\newblock


\bibitem[\protect\citeauthoryear{Blumenthal and Wijnen}{Blumenthal and
  Wijnen}{2002}]%
        {IETF-RFC3414}
\bibfield{author}{\bibinfo{person}{U. Blumenthal} {and} \bibinfo{person}{B.
  Wijnen}.} \bibinfo{year}{2002}\natexlab{}.
\newblock \bibinfo{title}{{User-based Security Model (USM) for version 3 of the
  Simple Network Management Protocol (SNMPv3)}}.
\newblock \bibinfo{howpublished}{{IETF RFC 3414}}.
\newblock


\bibitem[\protect\citeauthoryear{{CAIDA}}{{CAIDA}}{2021}]%
        {itdk}
\bibfield{author}{\bibinfo{person}{{CAIDA}}.} \bibinfo{year}{2021}\natexlab{}.
\newblock \bibinfo{title}{{Macroscopic Internet Topology Data Kit (ITDK)}}.
\newblock
  \bibinfo{howpublished}{\url{https://www.caida.org/catalog/datasets/internet-topology-data-kit/}}.
\newblock


\bibitem[\protect\citeauthoryear{CAIDA}{CAIDA}{2021}]%
        {caidaas-rank}
\bibfield{author}{\bibinfo{person}{CAIDA}.} \bibinfo{year}{2021}\natexlab{}.
\newblock \bibinfo{title}{{The CAIDA AS Ranking}}.
\newblock \bibinfo{howpublished}{\url{https://asrank.caida.org/}}.
\newblock


\bibitem[\protect\citeauthoryear{Case, Fedor, Schoffstall, and Davin}{Case
  et~al\mbox{.}}{1988}]%
        {IETF-RFC1067}
\bibfield{author}{\bibinfo{person}{J. Case}, \bibinfo{person}{M. Fedor},
  \bibinfo{person}{M. Schoffstall}, {and} \bibinfo{person}{J. Davin}.}
  \bibinfo{year}{1988}\natexlab{}.
\newblock \bibinfo{title}{{A Simple Network Management Protocol}}.
\newblock \bibinfo{howpublished}{{IETF RFC 1067}}.
\newblock


\bibitem[\protect\citeauthoryear{Case, Fedor, Schoffstall, and Davin}{Case
  et~al\mbox{.}}{1990}]%
        {IETF-RFC1157}
\bibfield{author}{\bibinfo{person}{J. Case}, \bibinfo{person}{M. Fedor},
  \bibinfo{person}{M. Schoffstall}, {and} \bibinfo{person}{J. Davin}.}
  \bibinfo{year}{1990}\natexlab{}.
\newblock \bibinfo{title}{{A Simple Network Management Protocol (SNMP)}}.
\newblock \bibinfo{howpublished}{{IETF RFC 1157}}.
\newblock


\bibitem[\protect\citeauthoryear{Case, McCloghrie, Rose, and Waldbusser}{Case
  et~al\mbox{.}}{1993a}]%
        {IETF-RFC1452}
\bibfield{author}{\bibinfo{person}{J. Case}, \bibinfo{person}{K. McCloghrie},
  \bibinfo{person}{M. Rose}, {and} \bibinfo{person}{S. Waldbusser}.}
  \bibinfo{year}{1993}\natexlab{a}.
\newblock \bibinfo{title}{{Coexistence between version 1 and version 2 of the
  Internet-standard Network Management Framework}}.
\newblock \bibinfo{howpublished}{{IETF RFC 1452}}.
\newblock


\bibitem[\protect\citeauthoryear{Case, McCloghrie, Rose, and Waldbusser}{Case
  et~al\mbox{.}}{1993b}]%
        {IETF-RFC1441}
\bibfield{author}{\bibinfo{person}{J. Case}, \bibinfo{person}{K. McCloghrie},
  \bibinfo{person}{M. Rose}, {and} \bibinfo{person}{S. Waldbusser}.}
  \bibinfo{year}{1993}\natexlab{b}.
\newblock \bibinfo{title}{{Introduction to version 2 of the Internet-standard
  Network Management Framework}}.
\newblock \bibinfo{howpublished}{{IETF RFC 1441}}.
\newblock


\bibitem[\protect\citeauthoryear{Case, McCloghrie, Rose, and Waldbusser}{Case
  et~al\mbox{.}}{1996a}]%
        {IETF-RFC1908}
\bibfield{author}{\bibinfo{person}{J. Case}, \bibinfo{person}{K. McCloghrie},
  \bibinfo{person}{M. Rose}, {and} \bibinfo{person}{S. Waldbusser}.}
  \bibinfo{year}{1996}\natexlab{a}.
\newblock \bibinfo{title}{{Coexistence between Version 1 and Version 2 of the
  Internet-standard Network Management Framework}}.
\newblock \bibinfo{howpublished}{{IETF RFC 1908}}.
\newblock


\bibitem[\protect\citeauthoryear{Case, McCloghrie, Rose, and Waldbusser}{Case
  et~al\mbox{.}}{1996b}]%
        {IETF-RFC1901}
\bibfield{author}{\bibinfo{person}{J. Case}, \bibinfo{person}{K. McCloghrie},
  \bibinfo{person}{M. Rose}, {and} \bibinfo{person}{S. Waldbusser}.}
  \bibinfo{year}{1996}\natexlab{b}.
\newblock \bibinfo{title}{{Introduction to Community-based SNMPv2}}.
\newblock \bibinfo{howpublished}{{IETF RFC 1901}}.
\newblock


\bibitem[\protect\citeauthoryear{Censys}{Censys}{2021}]%
        {Censys}
\bibfield{author}{\bibinfo{person}{Censys}.} \bibinfo{year}{2021}\natexlab{}.
\newblock \bibinfo{title}{{Censys Scanning and Data Collection}}.
\newblock \bibinfo{howpublished}{\url{https://censys.io/}}.
\newblock


\bibitem[\protect\citeauthoryear{Cisco}{Cisco}{2021a}]%
        {Cisco-EngineID}
\bibfield{author}{\bibinfo{person}{Cisco}.} \bibinfo{year}{2021}\natexlab{a}.
\newblock \bibinfo{title}{{Catalyst 2960 and 2960-S Software Configuration
  Guide, 12.2(55)SE}}.
\newblock
  \bibinfo{howpublished}{\url{https://www.cisco.com/c/en/us/td/docs/switches/lan/catalyst2960/software/release/12-2_55_se/configuration/guide/scg_2960/swsnmp.html}}.
\newblock


\bibitem[\protect\citeauthoryear{Cisco}{Cisco}{2021b}]%
        {cisco-bug-engineid}
\bibfield{author}{\bibinfo{person}{Cisco}.} \bibinfo{year}{2021}\natexlab{b}.
\newblock \bibinfo{title}{{Cisco Bug: CSCts87275 - Cat4k with sup7e : same snmp
  engineID on different cat4k switches}}.
\newblock
  \bibinfo{howpublished}{\url{https://quickview.cloudapps.cisco.com/quickview/bug/CSCts87275}}.
\newblock


\bibitem[\protect\citeauthoryear{{Cisco PSIRT}}{{Cisco PSIRT}}{2021a}]%
        {cisco00727}
\bibfield{author}{\bibinfo{person}{{Cisco PSIRT}}.}
  \bibinfo{year}{2021}\natexlab{a}.
\newblock \bibinfo{title}{{SNMP v3 information leaking vulnerability}}.
\newblock
\newblock
\newblock
\shownote{\url{https://bst.cloudapps.cisco.com/bugsearch/bug/CSCuz00727}.}


\bibitem[\protect\citeauthoryear{{Cisco PSIRT}}{{Cisco PSIRT}}{2021b}]%
        {cisco74132}
\bibfield{author}{\bibinfo{person}{{Cisco PSIRT}}.}
  \bibinfo{year}{2021}\natexlab{b}.
\newblock \bibinfo{title}{{SNMP v3 information leaking vulnerability}}.
\newblock
\newblock
\newblock
\shownote{\url{https://bst.cloudapps.cisco.com/bugsearch/bug/CSCtw74132}.}


\bibitem[\protect\citeauthoryear{Czyz, Luckie, Allman, and Bailey}{Czyz
  et~al\mbox{.}}{2016}]%
        {Back-Door-IPv6}
\bibfield{author}{\bibinfo{person}{J. Czyz}, \bibinfo{person}{M. Luckie},
  \bibinfo{person}{M. Allman}, {and} \bibinfo{person}{M. Bailey}.}
  \bibinfo{year}{2016}\natexlab{}.
\newblock \showarticletitle{{Don't Forget to Lock the Back Door! A
  Characterization of IPv6 Network Security Policy}}. In
  \bibinfo{booktitle}{\emph{NDSS}}.
\newblock


\bibitem[\protect\citeauthoryear{Dittrich, Kenneally, et~al\mbox{.}}{Dittrich
  et~al\mbox{.}}{2012}]%
        {dittrich2012menlo}
\bibfield{author}{\bibinfo{person}{D. Dittrich}, \bibinfo{person}{E.
  Kenneally}, {et~al\mbox{.}}} \bibinfo{year}{2012}\natexlab{}.
\newblock \showarticletitle{{The Menlo Report: Ethical Principles Guiding
  Information and Communication Technology Research}}.
\newblock \bibinfo{journal}{\emph{U.S. Department of Homeland Security}}
  (\bibinfo{year}{2012}).
\newblock


\bibitem[\protect\citeauthoryear{Durumeric, Adrian, Mirian, Bailey, and
  Halderman}{Durumeric et~al\mbox{.}}{2015}]%
        {search-engine-CCS2015}
\bibfield{author}{\bibinfo{person}{Z. Durumeric}, \bibinfo{person}{D. Adrian},
  \bibinfo{person}{A. Mirian}, \bibinfo{person}{M. Bailey}, {and}
  \bibinfo{person}{J.~A. Halderman}.} \bibinfo{year}{2015}\natexlab{}.
\newblock \showarticletitle{{A Search Engine Backed by Internet-Wide
  Scanning}}. In \bibinfo{booktitle}{\emph{ACM CCS}}.
\newblock


\bibitem[\protect\citeauthoryear{Durumeric, Wustrow, and Halderman}{Durumeric
  et~al\mbox{.}}{2013}]%
        {ZMap}
\bibfield{author}{\bibinfo{person}{Z. Durumeric}, \bibinfo{person}{E. Wustrow},
  {and} \bibinfo{person}{J.~A. Halderman}.} \bibinfo{year}{2013}\natexlab{}.
\newblock \showarticletitle{{ZMap: Fast Internet-Wide Scanning and its Security
  Applications}}. In \bibinfo{booktitle}{\emph{USENIX Security Symposium}}.
\newblock


\bibitem[\protect\citeauthoryear{Gasser, Scheitle, Foremski, Lone, Korczynski,
  Strowes, Hendriks, and Carle}{Gasser et~al\mbox{.}}{2018}]%
        {gasser2018clusters}
\bibfield{author}{\bibinfo{person}{O. Gasser}, \bibinfo{person}{Q. Scheitle},
  \bibinfo{person}{P. Foremski}, \bibinfo{person}{Q. Lone}, \bibinfo{person}{M.
  Korczynski}, \bibinfo{person}{S.~D. Strowes}, \bibinfo{person}{L. Hendriks},
  {and} \bibinfo{person}{G. Carle}.} \bibinfo{year}{2018}\natexlab{}.
\newblock \showarticletitle{{Clusters in the Expanse: Understanding and
  Unbiasing IPv6 Hitlists}}. In \bibinfo{booktitle}{\emph{ACM IMC}}.
\newblock


\bibitem[\protect\citeauthoryear{Gasser, Scheitle, Foremski, Lone, Korczynski,
  Strowes, Hendriks, and Carle}{Gasser et~al\mbox{.}}{2021}]%
        {v6hl}
\bibfield{author}{\bibinfo{person}{O. Gasser}, \bibinfo{person}{Q. Scheitle},
  \bibinfo{person}{P. Foremski}, \bibinfo{person}{Q. Lone}, \bibinfo{person}{M.
  Korczynski}, \bibinfo{person}{S.~D. Strowes}, \bibinfo{person}{L. Hendriks},
  {and} \bibinfo{person}{G. Carle}.} \bibinfo{year}{2021}\natexlab{}.
\newblock \bibinfo{title}{{IPv6 Hitlist Service}}.
\newblock \bibinfo{howpublished}{\url{https://ipv6hitlist.github.io/}}.
\newblock


\bibitem[\protect\citeauthoryear{Gasser, Scheitle, Rudolph, Denis, Schricker,
  and Carle}{Gasser et~al\mbox{.}}{2017}]%
        {gasser2017amplification}
\bibfield{author}{\bibinfo{person}{O. Gasser}, \bibinfo{person}{Q. Scheitle},
  \bibinfo{person}{B. Rudolph}, \bibinfo{person}{C. Denis}, \bibinfo{person}{N.
  Schricker}, {and} \bibinfo{person}{G. Carle}.}
  \bibinfo{year}{2017}\natexlab{}.
\newblock \showarticletitle{{The Amplification Threat Posed by Publicly
  Reachable BACnet Devices}}.
\newblock \bibinfo{journal}{\emph{Journal of Cyber Security and Mobility}}
  \bibinfo{volume}{6} (\bibinfo{year}{2017}).
\newblock
Issue 1.


\bibitem[\protect\citeauthoryear{Govindan and Tangmunarunkit}{Govindan and
  Tangmunarunkit}{2000}]%
        {govindan2000heuristics}
\bibfield{author}{\bibinfo{person}{R. Govindan} {and} \bibinfo{person}{H.
  Tangmunarunkit}.} \bibinfo{year}{2000}\natexlab{}.
\newblock \showarticletitle{{Heuristics for Internet Map Discovery}}. In
  \bibinfo{booktitle}{\emph{IEEE INFOCOM}}.
\newblock


\bibitem[\protect\citeauthoryear{Gunes and Sarac}{Gunes and Sarac}{2006}]%
        {gunes2006analytical}
\bibfield{author}{\bibinfo{person}{M.~H. Gunes} {and} \bibinfo{person}{K.
  Sarac}.} \bibinfo{year}{2006}\natexlab{}.
\newblock \showarticletitle{{Analytical IP alias resolution}}. In
  \bibinfo{booktitle}{\emph{2006 IEEE ICC}}.
\newblock


\bibitem[\protect\citeauthoryear{Hardaker}{Hardaker}{2011}]%
        {rfc6353}
\bibfield{author}{\bibinfo{person}{W. Hardaker}.}
  \bibinfo{year}{2011}\natexlab{}.
\newblock \bibinfo{title}{{Transport Layer Security (TLS) Transport Model for
  the Simple Network Management Protocol (SNMP)}}.
\newblock \bibinfo{howpublished}{RFC 6353 (Internet Standard)}.
\newblock
\showISSN{2070-1721}
\urldef\tempurl%
\url{https://doi.org/10.17487/RFC6353}
\showDOI{\tempurl}


\bibitem[\protect\citeauthoryear{Harrington, Presuhn, and Wijnen}{Harrington
  et~al\mbox{.}}{2002}]%
        {IETF-RFC3411}
\bibfield{author}{\bibinfo{person}{D. Harrington}, \bibinfo{person}{R.
  Presuhn}, {and} \bibinfo{person}{B. Wijnen}.}
  \bibinfo{year}{2002}\natexlab{}.
\newblock \bibinfo{title}{{An Architecture for Describing Simple Network
  Management Protocol (SNMP) Management Frameworks}}.
\newblock \bibinfo{howpublished}{{IETF RFC 3411}}.
\newblock


\bibitem[\protect\citeauthoryear{Holland, Teixeira, Schmitt, Borgolte, Rexford,
  Feamster, and Mayer}{Holland et~al\mbox{.}}{2020}]%
        {Classifying-Vendors}
\bibfield{author}{\bibinfo{person}{J. Holland}, \bibinfo{person}{R. Teixeira},
  \bibinfo{person}{P. Schmitt}, \bibinfo{person}{K. Borgolte},
  \bibinfo{person}{J. Rexford}, \bibinfo{person}{N. Feamster}, {and}
  \bibinfo{person}{J. Mayer}.} \bibinfo{year}{2020}\natexlab{}.
\newblock \bibinfo{title}{{Classifying Network Vendors at Internet Scale}}.
\newblock \bibinfo{howpublished}{arXiv,
  \url{https://arxiv.org/abs/2006.13086}}.
\newblock


\bibitem[\protect\citeauthoryear{Huawei}{Huawei}{2021}]%
        {Huawei-EngineID}
\bibfield{author}{\bibinfo{person}{Huawei}.} \bibinfo{year}{2021}\natexlab{}.
\newblock \bibinfo{title}{{S2750, S5700, S6700 V200R003(C00, C02, and C10)
  Configuration Guide - Network Management and Monitoring}}.
\newblock
  \bibinfo{howpublished}{\url{https://support.huawei.com/enterprise/en/doc/EDOC1000027472?section=j005}}.
\newblock


\bibitem[\protect\citeauthoryear{{IEEE}}{{IEEE}}{2021}]%
        {macoui}
\bibfield{author}{\bibinfo{person}{{IEEE}}.} \bibinfo{year}{2021}\natexlab{}.
\newblock \bibinfo{title}{{List of MAC OUIs}}.
\newblock
  \bibinfo{howpublished}{\url{http://standards-oui.ieee.org/oui/oui.txt}}.
\newblock


\bibitem[\protect\citeauthoryear{Izhikevich, Teixeira, and
  Durumeric}{Izhikevich et~al\mbox{.}}{2021}]%
        {LZR2021}
\bibfield{author}{\bibinfo{person}{L. Izhikevich}, \bibinfo{person}{R.
  Teixeira}, {and} \bibinfo{person}{Z. Durumeric}.}
  \bibinfo{year}{2021}\natexlab{}.
\newblock \showarticletitle{{LZR: Identifying Unexpected Internet Services}}.
  In \bibinfo{booktitle}{\emph{{USENIX} Security Symposium}}.
\newblock


\bibitem[\protect\citeauthoryear{Keys, Hyun, Luckie, and k~claffy}{Keys
  et~al\mbox{.}}{2013}]%
        {keys13midar}
\bibfield{author}{\bibinfo{person}{K. Keys}, \bibinfo{person}{Y. Hyun},
  \bibinfo{person}{M. Luckie}, {and} \bibinfo{person}{k claffy}.}
  \bibinfo{year}{2013}\natexlab{}.
\newblock \showarticletitle{{Internet-Scale IPv4 Alias Resolution with MIDAR}}.
\newblock \bibinfo{journal}{\emph{IEEE/ACM Trans. Networking}}
  \bibinfo{volume}{21}, \bibinfo{number}{2} (\bibinfo{year}{2013}).
\newblock


\bibitem[\protect\citeauthoryear{Kohno, Broido, and Claffy}{Kohno
  et~al\mbox{.}}{2005}]%
        {kohno2005remote}
\bibfield{author}{\bibinfo{person}{T. Kohno}, \bibinfo{person}{A. Broido},
  {and} \bibinfo{person}{KC Claffy}.} \bibinfo{year}{2005}\natexlab{}.
\newblock \showarticletitle{{Remote Physical Device Fingerprinting}}.
\newblock \bibinfo{journal}{\emph{IEEE Transactions on Dependable and Secure
  Computing}} \bibinfo{volume}{2}, \bibinfo{number}{2} (\bibinfo{year}{2005}).
\newblock


\bibitem[\protect\citeauthoryear{K\"uhrer, Hupperich, Rossow, and
  Holz}{K\"uhrer et~al\mbox{.}}{2014}]%
        {Hell-Handshake}
\bibfield{author}{\bibinfo{person}{M. K\"uhrer}, \bibinfo{person}{T.
  Hupperich}, \bibinfo{person}{C. Rossow}, {and} \bibinfo{person}{T. Holz}.}
  \bibinfo{year}{2014}\natexlab{}.
\newblock \showarticletitle{{Hell of a Handshake: Abusing TCP for Reflective
  Amplification DDoS Attacks}}. In \bibinfo{booktitle}{\emph{{USENIX} WOOT}}.
\newblock


\bibitem[\protect\citeauthoryear{Luckie and Beverly}{Luckie and
  Beverly}{2017}]%
        {Router-outages-SIGCOMM2017}
\bibfield{author}{\bibinfo{person}{M. Luckie} {and} \bibinfo{person}{R.
  Beverly}.} \bibinfo{year}{2017}\natexlab{}.
\newblock \showarticletitle{{The Impact of Router Outages on the AS-level
  Internet}}. In \bibinfo{booktitle}{\emph{ACM SIGCOMM}}.
\newblock


\bibitem[\protect\citeauthoryear{Luckie, Beverly, and Brinkmeyer}{Luckie
  et~al\mbox{.}}{2013}]%
        {Speedtrap}
\bibfield{author}{\bibinfo{person}{M. Luckie}, \bibinfo{person}{R. Beverly},
  {and} \bibinfo{person}{W. Brinkmeyer}.} \bibinfo{year}{2013}\natexlab{}.
\newblock \showarticletitle{{Speedtrap: Internet-Scale IPv6 Alias Resolution}}.
  In \bibinfo{booktitle}{\emph{ACM IMC}}.
\newblock


\bibitem[\protect\citeauthoryear{Luckie, Huffaker, and k.~claffy}{Luckie
  et~al\mbox{.}}{2019}]%
        {regex19}
\bibfield{author}{\bibinfo{person}{M. Luckie}, \bibinfo{person}{B. Huffaker},
  {and} \bibinfo{person}{k. claffy}.} \bibinfo{year}{2019}\natexlab{}.
\newblock \showarticletitle{{Learning Regexes to Extract Router Names from
  Hostnames}}. In \bibinfo{booktitle}{\emph{ACM IMC}}.
\newblock


\bibitem[\protect\citeauthoryear{Marder}{Marder}{2020}]%
        {marder2020apple}
\bibfield{author}{\bibinfo{person}{A. Marder}.}
  \bibinfo{year}{2020}\natexlab{}.
\newblock \showarticletitle{{APPLE: Alias Pruning by Path Length Estimation}}.
  In \bibinfo{booktitle}{\emph{PAM}}.
\newblock


\bibitem[\protect\citeauthoryear{MITRE}{MITRE}{2021}]%
        {snmp-cve}
\bibfield{author}{\bibinfo{person}{MITRE}.} \bibinfo{year}{2021}\natexlab{}.
\newblock \bibinfo{title}{{Common Vulnerabilities and Exposures, SNMP}}.
\newblock
  \bibinfo{howpublished}{\url{https://cve.mitre.org/cgi-bin/cvekey.cgi?keyword=SNMP}}.
\newblock


\bibitem[\protect\citeauthoryear{Moore}{Moore}{2020}]%
        {hdmoore}
\bibfield{author}{\bibinfo{person}{HD Moore}.} \bibinfo{year}{2020}\natexlab{}.
\newblock \bibinfo{title}{{Security Surprises with SNMP v3}}.
\newblock
\newblock
\newblock
\shownote{\url{https://www.rumble.run/blog/security-surprises-with-snmp-v3/}.}


\bibitem[\protect\citeauthoryear{Murdoch}{Murdoch}{2006}]%
        {murdoch2006hot}
\bibfield{author}{\bibinfo{person}{S.~J. Murdoch}.}
  \bibinfo{year}{2006}\natexlab{}.
\newblock \showarticletitle{{Hot or not: Revealing hidden services by their
  clock skew}}. In \bibinfo{booktitle}{\emph{ACM CCS}}.
\newblock


\bibitem[\protect\citeauthoryear{NCC}{NCC}{2021}]%
        {RIPE-Atlas}
\bibfield{author}{\bibinfo{person}{RIPE NCC}.} \bibinfo{year}{2021}\natexlab{}.
\newblock \bibinfo{title}{{RIPE Atlas}}.
\newblock \bibinfo{howpublished}{\url{https://atlas.ripe.net/}}.
\newblock


\bibitem[\protect\citeauthoryear{{Net-SNMP Project}}{{Net-SNMP
  Project}}{2021}]%
        {netsnmp}
\bibfield{author}{\bibinfo{person}{{Net-SNMP Project}}.}
  \bibinfo{year}{2021}\natexlab{}.
\newblock \bibinfo{title}{{Net-SNMP}}.
\newblock \bibinfo{howpublished}{\url{http://www.net-snmp.org/}}.
\newblock


\bibitem[\protect\citeauthoryear{Nmap}{Nmap}{2021}]%
        {Nmap}
\bibfield{author}{\bibinfo{person}{Nmap}.} \bibinfo{year}{2021}\natexlab{}.
\newblock \bibinfo{title}{{Nmap: the Network Mapper - Free Security Scanner}}.
\newblock \bibinfo{howpublished}{\url{https://nmap.org/}}.
\newblock


\bibitem[\protect\citeauthoryear{Partridge and Allman}{Partridge and
  Allman}{2016}]%
        {partridge2016ethical}
\bibfield{author}{\bibinfo{person}{C. Partridge} {and} \bibinfo{person}{M.
  Allman}.} \bibinfo{year}{2016}\natexlab{}.
\newblock \showarticletitle{{Ethical Considerations in Network Measurement
  Papers}}.
\newblock \bibinfo{journal}{\emph{Commun. ACM}} (\bibinfo{year}{2016}).
\newblock


\bibitem[\protect\citeauthoryear{Rossow}{Rossow}{2014}]%
        {rossow2014amplification}
\bibfield{author}{\bibinfo{person}{C. Rossow}.}
  \bibinfo{year}{2014}\natexlab{}.
\newblock \showarticletitle{{Amplification Hell: Revisiting Network Protocols
  for DDoS Abuse}}. In \bibinfo{booktitle}{\emph{NDSS}}.
\newblock


\bibitem[\protect\citeauthoryear{Rye and Beverly}{Rye and Beverly}{2020}]%
        {pam20edgy}
\bibfield{author}{\bibinfo{person}{E.~C. Rye} {and} \bibinfo{person}{R.
  Beverly}.} \bibinfo{year}{2020}\natexlab{}.
\newblock \showarticletitle{{Discovering the IPv6 Network Periphery}}. In
  \bibinfo{booktitle}{\emph{PAM}}.
\newblock


\bibitem[\protect\citeauthoryear{Rye, Beverly, and kc~claffy}{Rye
  et~al\mbox{.}}{2021}]%
        {imc21scent}
\bibfield{author}{\bibinfo{person}{E.~C. Rye}, \bibinfo{person}{R. Beverly},
  {and} \bibinfo{person}{kc claffy}.} \bibinfo{year}{2021}\natexlab{}.
\newblock \showarticletitle{{Follow the Scent: Defeating IPv6 Prefix Rotation
  Privacy}}. In \bibinfo{booktitle}{\emph{ACM IMC}}.
\newblock


\bibitem[\protect\citeauthoryear{Sargent, Kristoff, Paxson, and Allman}{Sargent
  et~al\mbox{.}}{2017}]%
        {sargent2017potential}
\bibfield{author}{\bibinfo{person}{M. Sargent}, \bibinfo{person}{J. Kristoff},
  \bibinfo{person}{V. Paxson}, {and} \bibinfo{person}{M. Allman}.}
  \bibinfo{year}{2017}\natexlab{}.
\newblock \showarticletitle{{On the Potential Abuse of IGMP}}.
\newblock \bibinfo{journal}{\emph{ACM CCR}} (\bibinfo{year}{2017}).
\newblock


\bibitem[\protect\citeauthoryear{Scheitle, Gasser, Rouhi, and Carle}{Scheitle
  et~al\mbox{.}}{2017}]%
        {IPv4-6-Siblings}
\bibfield{author}{\bibinfo{person}{Q. Scheitle}, \bibinfo{person}{O. Gasser},
  \bibinfo{person}{M. Rouhi}, {and} \bibinfo{person}{G. Carle}.}
  \bibinfo{year}{2017}\natexlab{}.
\newblock \showarticletitle{{Large-Scale Classification of IPv6-IPv4 Siblings
  with Variable Clock Skew}}. In \bibinfo{booktitle}{\emph{TMA}}.
\newblock


\bibitem[\protect\citeauthoryear{{Shadowserver}}{{Shadowserver}}{2021}]%
        {shadowserver-snmp}
\bibfield{author}{\bibinfo{person}{{Shadowserver}}.}
  \bibinfo{year}{2021}\natexlab{}.
\newblock \bibinfo{title}{{Open SNMP Scanning Project}}.
\newblock \bibinfo{howpublished}{\url{https://scan.shadowserver.org/snmp/}}.
\newblock


\bibitem[\protect\citeauthoryear{Sherry, Katz-Bassett, Pimenova, Madhyastha,
  Anderson, and Krishnamurthy}{Sherry et~al\mbox{.}}{2010}]%
        {sherry2010resolving}
\bibfield{author}{\bibinfo{person}{J. Sherry}, \bibinfo{person}{E.
  Katz-Bassett}, \bibinfo{person}{M. Pimenova}, \bibinfo{person}{H.
  Madhyastha}, \bibinfo{person}{T. Anderson}, {and} \bibinfo{person}{A.
  Krishnamurthy}.} \bibinfo{year}{2010}\natexlab{}.
\newblock \showarticletitle{{Resolving IP aliases with prespecified
  timestamps}}. In \bibinfo{booktitle}{\emph{ACM IMC}}.
\newblock


\bibitem[\protect\citeauthoryear{Spring, Mahajan, and Wetherall}{Spring
  et~al\mbox{.}}{2002}]%
        {Rocketfuel}
\bibfield{author}{\bibinfo{person}{N. Spring}, \bibinfo{person}{R. Mahajan},
  {and} \bibinfo{person}{D. Wetherall}.} \bibinfo{year}{2002}\natexlab{}.
\newblock \showarticletitle{{Measuring ISP topologies with Rocketfuel}}. In
  \bibinfo{booktitle}{\emph{ACM SIGCOMM}}.
\newblock


\bibitem[\protect\citeauthoryear{Stallings}{Stallings}{1998}]%
        {stallings}
\bibfield{author}{\bibinfo{person}{W. Stallings}.}
  \bibinfo{year}{1998}\natexlab{}.
\newblock \showarticletitle{{SNMPv3: A security enhancement for SNMP}}.
\newblock \bibinfo{journal}{\emph{IEEE Communications Surveys}}
  \bibinfo{volume}{1}, \bibinfo{number}{1} (\bibinfo{year}{1998}),
  \bibinfo{pages}{2--17}.
\newblock


\bibitem[\protect\citeauthoryear{{The Mitre Corporation}}{{The Mitre
  Corporation}}{2012}]%
        {cve}
\bibfield{author}{\bibinfo{person}{{The Mitre Corporation}}.}
  \bibinfo{year}{2012}\natexlab{}.
\newblock \bibinfo{title}{{CVE-2012-5719}}.
\newblock
\newblock
\newblock
\shownote{\url{https://cve.mitre.org/cgi-bin/cvename.cgi?name=2012-5719}.}


\bibitem[\protect\citeauthoryear{Thomas}{Thomas}{2021}]%
        {SNMPv3-bruteforce-attacks}
\bibfield{author}{\bibinfo{person}{S. Thomas}.}
  \bibinfo{year}{2021}\natexlab{}.
\newblock \bibinfo{title}{{Brute forcing SNMPv3 authentication}}.
\newblock
  \bibinfo{howpublished}{\url{https://applied-risk.com/resources/brute-forcing-snmpv3-authentication}}.
\newblock


\bibitem[\protect\citeauthoryear{Vanaubel, Pansiot, Merindol, and
  Donnet}{Vanaubel et~al\mbox{.}}{2013}]%
        {TTL-fingerprinting}
\bibfield{author}{\bibinfo{person}{Y. Vanaubel}, \bibinfo{person}{J.-J.
  Pansiot}, \bibinfo{person}{P. Merindol}, {and} \bibinfo{person}{B. Donnet}.}
  \bibinfo{year}{2013}\natexlab{}.
\newblock \showarticletitle{{Network Fingerprinting: TTL-Based Router
  Signatures}}. In \bibinfo{booktitle}{\emph{ACM IMC}}.
\newblock


\bibitem[\protect\citeauthoryear{Vermeulen, Ljuma, Addanki, Gouel, Fourmaux,
  Friedman, and Rejaie}{Vermeulen et~al\mbox{.}}{2020}]%
        {vermeulen2020alias}
\bibfield{author}{\bibinfo{person}{K. Vermeulen}, \bibinfo{person}{B. Ljuma},
  \bibinfo{person}{V. Addanki}, \bibinfo{person}{M. Gouel}, \bibinfo{person}{O.
  Fourmaux}, \bibinfo{person}{T. Friedman}, {and} \bibinfo{person}{R. Rejaie}.}
  \bibinfo{year}{2020}\natexlab{}.
\newblock \showarticletitle{{Alias Resolution Based on ICMP Rate Limiting}}. In
  \bibinfo{booktitle}{\emph{PAM}}.
\newblock


\bibitem[\protect\citeauthoryear{Zander and Murdoch}{Zander and
  Murdoch}{2008}]%
        {zander2008improved}
\bibfield{author}{\bibinfo{person}{S. Zander} {and} \bibinfo{person}{S.~J.
  Murdoch}.} \bibinfo{year}{2008}\natexlab{}.
\newblock \showarticletitle{{An Improved Clock-skew Measurement Technique for
  Revealing Hidden Services}}. In \bibinfo{booktitle}{\emph{USENIX Security
  Symposium}}.
\newblock


\end{thebibliography}

\begin{table*}[!bpt]
{%
\begin{tabular}{lrrrr}
\toprule
{} & Alias sets & Non-singleton alias sets & IPs in non-singleton alias sets &  IPs per non-singleton alias set \\
\midrule
Exact first              &       5.3M &                     903k &                            8.2M &                              9.1 \\
Exact both               &       5.9M &                     892k &                            7.5M &                              8.4 \\
Round first              &       4.6M &                     826k &                            8.7M &                             10.6 \\
Round both               &       4.7M &                     835k &                            8.7M &                             10.4 \\
Divide by 20 first       &       4.6M &                     820k &                            8.8M &                             10.7 \\
Divide by 20 both        &       4.6M &                     824k &                            8.7M &                             10.6 \\
Divide by 20+round first &       4.6M &                     820k &                            8.8M &                             10.7 \\
Divide by 20+round both  &       4.6M &                     824k &                            8.7M &                             10.6 \\
\bottomrule
\end{tabular}
}
\caption{Comparison of different alias resolution approaches.}
\label{table:aliasresolution}
\end{table*}

\appendix
\section{Comparison of different alias resolution approaches}\label{sec:appendix:aliasresolution}

\Cref{table:aliasresolution} shows the results for different alias resolution
variations.  ``First'' and ``both'' revers to using fields from the first scan
only vs. using fields from both scans for the matching process.  The techniques
differ only on the matching threshold for the \lastreboot.  All other fields are
matched exactly.  ``Exact'' denotes an exact matching of \lastreboot, ``Round''
means that the last digit is rounded, ``Divide by 20'' means that the last digit
is divided by 20 and cut off (\ie put into bins of 20 seconds), ``Divide by
20+round'' denotes division by 20 and rounding of the resulting floating point
number.

\section{Uniqueness of \lastreboot and \etime}\label{sec:appendix:uniqueness}

\begin{figure}[t]
    \centering
    \includegraphics[width=\linewidth]{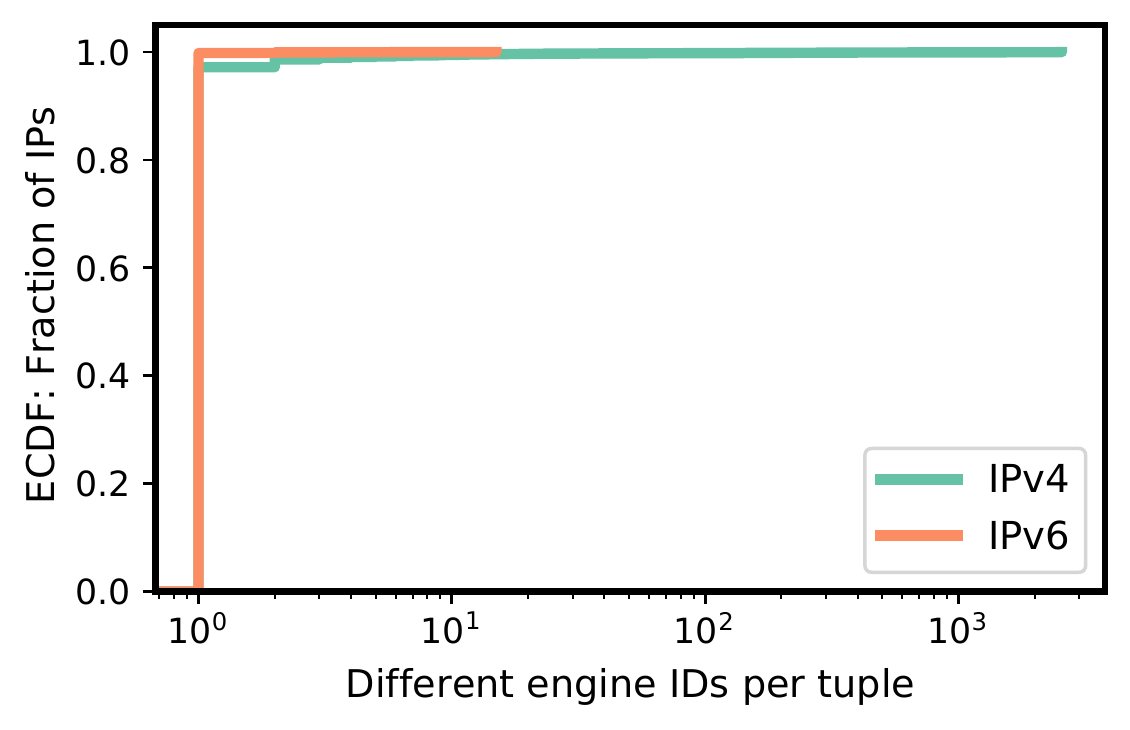}%
    \caption{Uniqueness of (\lastreboot, \eboots) tuples across IPv4 and IPv6 addresses: For the vast majority of tuples we see a single \eid. Note that the x axis is log-scaled.}
    \label{fig:lastreboot_boots_uniqueness_per_ip}
\end{figure}

In addition to the \eid SNMP send \etime and \eboots values in their response.
Our alias resolution technique leverages these information as well.
By subtracting the \etime from the current time of the scan we calculate a \lastreboot.
In order to learn more about the uniqueness of \lastreboot and \etime tuples, we look for cases where the same \lastreboot and \etime values are seen at different devices, \ie devices with different \eids.
\Cref{fig:lastreboot_boots_uniqueness_per_ip} shows the distribution of the same \lastreboot and \etime tuples for different number of engine IDs.
We find that for 97.2\% and 99.8\% of IPv4 and IPv6 addresses have \lastreboot and \etime tuples with a single unique \eid.
This shows that this tuple is indeed a valuable addition for our alias resolution technique.

\section{Distribution of Routers per AS and Region}\label{sec:dist-routers-AS-region}

\begin{figure}[t]
    \centering
    \includegraphics[width=\linewidth]{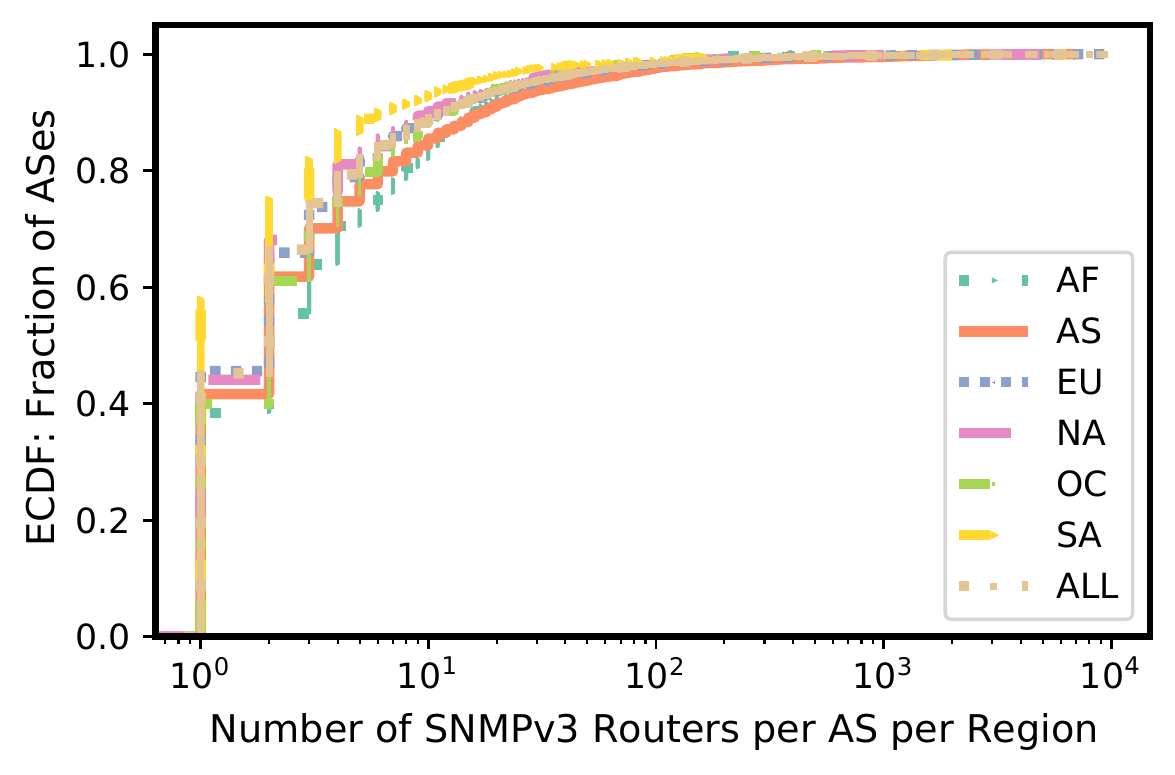}
\caption{Distribution of number of routers per AS for different regions.}
    \label{fig:router-coverage-all}
\end{figure}

In Figure~\ref{fig:router-coverage-all} we plot the distribution of routers that
our technique identified per network (AS) and per region. 
To map ASes to countries, we utilized CAIDA's AS Ranking~\cite{caidaas-rank}. This
results in an AS-to-country reaches 99,9\% for the ASes with routers our study,
\ie 22,769 out of 22,789. We are aware that a network may span multiple
countries within a continent, \ie we decided to map a network to a continent
(region). Again, there are networks that span multiple regions, but in this
case, typically, they register different ASes and address space in different
regions. If this is not the case then, we assign the region of the headquarter
of the network operator to the AS it owns. Our analysis does not show
significant distributional differences across continents. However, most of the
networks with largest number of routers are in North America (NA) and Europe
(EU).

\label{page:last}

\end{document}